\pdfoutput=1
\documentclass[% 
unsortedaddress,
 notitlepage,amsmath,amssymb,
 aps,
]{revtex4-1}
\usepackage[pdftex]{graphicx}% Include figure files
\usepackage{bm}% bold math
\usepackage{hyperref}
\usepackage{amsmath}
\usepackage[pdftex]{color}
\usepackage{subfigure}
\usepackage[utf8]{inputenc}
\usepackage[greek,english]{babel}
\usepackage{textgreek}

\newcommand{\textunderscript}[1]{$_{\text{#1}}$}
\renewcommand{\figurename}{\textbf{Figure}}
\begin{document}

\preprint{}
\title{Large Spin-to-Charge Conversion in Ultrathin Gold-Silicon Multilayers}

\author{Mohammed Salah El Hadri}
\affiliation {Center for Memory and Recording Research, University of California, San Diego, La Jolla, California, 92093-0401, USA}
\email{melhadri@ucsd.edu}

\author{Jonathan Gibbons}
\affiliation {Department of Physics, University of California San Diego, La Jolla, California, 92093-0401, USA}
\affiliation {Materials Science Division, Argonne National Laboratory, Argonne, Illinois 60439, USA}
\affiliation {Department of Materials Science and Engineering, University of Illinois at Urbana-Champaign, Urbana, Illinois 61801, USA}

\author{Yuxuan Xiao}
\affiliation {Center for Memory and Recording Research, University of California, San Diego, La Jolla, California, 92093-0401, USA}

\author{Haowen Ren}
\affiliation {Center for Memory and Recording Research, University of California, San Diego, La Jolla, California, 92093-0401, USA}

\author{Hanu Arava}
\affiliation {Northwestern-Argonne Institute of Science and Engineering (NAISE), Northwestern University, Evanston, Illinois 60208, USA}
\affiliation {Materials Science Division (MSD), Argonne National Laboratory, Argonne, Illinois 60439, USA}

\author{Yuzi Liu}
\affiliation {Center of Nanoscale Materials, Argonne National Laboratory, Argonne, Illinois 60439, USA}

\author{Zhaowei Liu}
\affiliation {Department of Electrical and Computer Engineering, University of California, San Diego, 9500 Gilman Drive, La Jolla, California 92093-0407, USA}
\affiliation {Materials Science and Engineering Program, University of California, San Diego, 9500 Gilman Drive, La Jolla, California 92093-0418, USA}
 
\author{Amanda Petford-Long}
\affiliation {Northwestern-Argonne Institute of Science and Engineering (NAISE), Northwestern University, Evanston, Illinois 60208, USA}
\affiliation {Materials Science Division (MSD), Argonne National Laboratory, Argonne, Illinois 60439, USA}

\author{Axel Hoffmann}
\affiliation {Materials Science Division, Argonne National Laboratory, Argonne, Illinois 60439, USA}
\affiliation {Department of Materials Science and Engineering, University of Illinois at Urbana-Champaign, Urbana, Illinois 61801, USA}

\author{Eric E. Fullerton}
\affiliation {Center for Memory and Recording Research, University of California, San Diego, La Jolla, California, 92093-0401, USA}
\affiliation {Department of Electrical and Computer Engineering, University of California, San Diego, 9500 Gilman Drive, La Jolla, California 92093-0407, USA}

\date{\today}% It is always \today, today,

\begin{abstract}

Investigation of the spin Hall effect in gold has triggered increasing interest over the past decade, since gold combines the properties of a large bulk spin diffusion length and strong interfacial spin-orbit coupling. However, discrepancies between the values of the spin Hall angle of gold reported in the literature have brought into question the microscopic origin of the spin Hall effect in Au. Here, we investigate the thickness dependence of the spin-charge conversion efficiency in single Au films and ultrathin Au/Si multilayers by non-local transport and spin-torque ferromagnetic resonance measurements. We show that the spin-charge conversion efficiency is strongly enhanced in ultrathin Au/Si multilayers, reaching exceedingly large values of 0.99 $\pm$ 0.34 when the thickness of the individual Au layers is scaled down to 2 nm. These findings reveal the coexistence of a strong interfacial spin-orbit coupling effect which becomes dominant in ultrathin Au, and bulk spin Hall effect with a relatively low bulk spin Hall angle of 0.012 $\pm$ 0.005. Our experimental results suggest the key role of the Rashba-Edelstein effect in the spin-to-charge conversion in ultrathin Au.

\end{abstract}
\maketitle
\section{Introduction}

The interplay between spin-orbit coupling (SOC) and low dimensionality has attracted significant interest over the last decade due to the prospect of exploring rich physical mechanisms, as well as the potential impact on emerging spintronics technologies \cite{Hellman2017,Soumyanarayanan2016,Manchon2015,Sklenar2016}. One of the most prominent mechanisms arising from the relativistic SOC is the spin Hall effect (SHE) \cite{Hoffmann2013,Sinova2015}, where a transverse spin current is generated from an unpolarized charge current. Since its initial prediction by D'yakonov and Perel in 1971 \cite{Dyakonov1971}, the spin Hall effect has attracted increasing interest from theoretical \cite{Hirsch1999,Zhang2000,Sinova2004} and experimental \cite{Valenzuela2006,Kato2004,Wunderlich2005,Saitoh2006,Kimura2007} viewpoints, and has become an important tool for the injection, detection and manipulation of spin currents in thin films and heterostructures \cite{Hoffmann2013,Sinova2015,Pai2012,Liu2012a,Liu2012b,Liu2012c,Manchon2019}. The microscopic origin of the SHE is a material-dependent combination of intrinsic and extrinsic mechanisms. Its efficiency is characterized by the spin Hall angle (SHA) (\textbf{$\theta$}\textunderscript{SHE}), which describes the conversion of charge into spin currents. Recent studies have explored new pathways to enhance the SHE by investigating novel chemically inhomogeneous \cite{Zhu2018,Demasius2016,An2018a,An2018b} and multilayer \cite{Zhu2019a,Zhu2019b} systems, and reported large spin-torque efficiencies, namely 0.35 for Au\textunderscript{1-x}Pt\textunderscript{x} alloys \cite{Zhu2018}, -0.49 for oxidized W \cite{Demasius2016}, 0.9 for oxidized Pt \cite{An2018a,An2018b}, 0.37 for Pt/Hf multilayers \cite{Zhu2019a}, and 0.35 for Pt/Ti multilayers \cite{Zhu2019b}. More importantly, the spin-torque efficiencies reported for these novel systems are much larger than the SHA values reported for single-element layers, namely -0.33 for \textbf{$\beta$}-W \cite{Pai2012}, $\sim$0.1 for Pt \cite{Althammer2013,Zhang2013} and 0.11 for Au \cite{Seki2008}, thus highlighting the importance of chemical inhomogeneity and heterostructures in the SHE mechanism. On the other hand, introducing low dimensionality to heavy metals with a large bulk SHE can also significantly enhance the spin-charge conversion efficiency via interfacial SOC-related effects, such as the Rashba SOC \cite{Rashba1960,Edelstein1990,Bychkov1984,Burkov2004}. Hence, these emergent interfacial SOC phenomena offer promising routes for 2D-spintronics applications \cite{Hellman2017,Soumyanarayanan2016,Manchon2015,Rojas-Sanchez2013,Jungfleisch2016}.

Since Au combines the properties of strong SOC at interfaces \cite{Seki2008,LaShell1996,Nicolay2001,Nechaev2009} and a relatively large bulk spin diffusion length \cite{Johnson1993,Ji2004,Ku2006}, it has been the focus of intense research interest, evidenced by many experimental and theoretical investigations. In 2008, Seki \textit{et al.} reported a giant SHA of 0.11 in 10-nm-thick Au using an FePt perpendicular spin injector \cite{Seki2008}. Later, non-local transport measurements by Mihajlovi\'{c} \textit{et al}. showed the absence of the giant SHE in a 60-nm-thick H-shaped Au structure (\textbf{$\theta$}\textunderscript{SHE} $\leq$ 0.023) \cite{Mihajlovic2009}, thus triggering a debate on the microscopic origin of the giant SHE in Au films. Follow up experimental investigations using various methods, such as non-local spin injection \cite{Seki2010}, spin pumping \cite{Brangham2016}, and non-local transport \cite{Tian2016,Chen2019}, have revealed that the SHA of Au can be significantly reduced for films thicker than 10 nm. More recently, first-principles calculations demonstrated the presence of a strong interfacial contribution to the SHE in Au-Fe bilayers, which was attributed to spin-dependent transmission occurring within a few atomic layers \cite{Li2019}. More importantly, such an interfacial contribution occurs at the Au(111) interface where a strong Rashba SOC exists \cite{LaShell1996,Nicolay2001,Nechaev2009}, which is a hint that the Rashba-splitting of the Au(111) surface states plays an important role in the spin-to-charge conversion mechanism in Au \cite{Li2019}. These investigations give insights into the origin of the large SHE in Au and suggest the important role of the interface scattering \cite{Seki2010,Brangham2016,Chen2019} and/or the Rashba SOC \cite{Li2019}.

To further elucidate the microscopic origin of the SHE in Au films, we present an experimental investigation of the SHE in ultrathin Au films and ultrathin Au/Si multilayers, where the thickness of the individual Au layers is scaled down to 2 nm. We utilized two different transport measurement techniques to probe the SHE, non-local transport and spin-torque ferromagnetic resonance (ST-FMR), which probe the spin currents flow in the ultrathin Au layers in the in-plane and the out-of-plane directions, respectively (see Fig. 1). In these two measurements, we are probing the spin-to-charge conversion in ultrathin Au arising from both bulk SHE and interfacial SOC-related effects. Hence, the spin-to-charge conversion in ultrathin Au will be characterized by a spin-charge conversion efficiency (\textbf{$\theta$}\textunderscript{sc}), rather than by a bulk SHA. We demonstrate that \textbf{$\theta$}\textunderscript{sc} is strongly enhanced in ultrathin Au/Si multilayers, reaching values much larger than all SHA values previously reported for Au films \cite{Seki2008,Brangham2016,Chen2019}. Moreover, a similar thickness-dependent behavior of \textbf{$\theta$}\textunderscript{sc} was obtained using the ST-FMR technique, however, with much lower \textbf{$\theta$}\textunderscript{sc} values. These findings indicate the coexistence of the bulk SHE with a relatively low bulk SHA and a strong interfacial SOC effect which becomes dominant in ultrathin Au. Our results rule out an interpretation of the thickness-dependent behavior of the spin-charge conversion efficiency in Au by the intrinsic SHE or the extrinsic skew scattering SHE mechanisms and, more importantly, suggest the key role of the Rashba-Edelstein mechanism in the giant spin-to-charge conversion in ultrathin Au.\\

\section{Sample fabrication} 
We have chosen to study the SHE in ultrathin Au/Si multilayers as they provide a unique opportunity to investigate spin-charge conversion efficiency of Au averaged over many individual Au layers, thereby giving more accuracy in the spin Hall properties measurements. Moreover, it has been shown experimentally that Au/Si(111) interfaces can have metallic spin-split surface states with an energy splitting up to 190 meV, which at the same time is very sensitive to the local structure \cite{Bondarenko2013}. Furthermore, Au/Si multilayers present a very promising system that exhibits the properties of hyperbolic metamaterials (HMMs), which have recently emerged as one of the prime candidates for extraordinary manipulation of light \cite{Shen2015,Qian2021}. In our non-local transport experiments, which probe the spin currents flow in the in-plane direction, as shown in Fig. 1(a), four different samples were investigated: two single polycrystalline Au films [Au (10 nm) and Au (60 nm)]; and two Au/Si multilayers [[Si (1.8 nm)/Au (2 nm)]\textunderscript{x5} and Au (5 nm)/[Si (4 nm)/Au (5 nm)]\textunderscript{x5}]. Each sample was deposited on a glass/Cr (3 nm) substrate using a DC sputtering technique. The Cr buffer layer is used to ensure good adhesion of the single Au layers and the Au/Si multilayers to the substrate (see Supplemental Material, Methods \cite{Supplemental}). The thickness of the insulating Si interlayers is chosen to be sufficiently large (\textit{t}\textunderscript{Si} $\geq$ 1.5 nm) to prevent strong coupling between the individual Au layers \cite{Cherradi1989}. To verify the multilayer nature of the studied sample with the thinnest Si and Au layer thicknesses of 1.8 nm and 2 nm, we performed dark-field cross-sectional scanning transmission electron microscopy (STEM). Fig. 1(c) shows the cross-sectional STEM view of the [Si (1.8 nm)/Au (2 nm)]\textunderscript{x5} multilayer structure, indicating a well-defined layered Au/Si structure and the absence of a strong cumulative roughness along the growth direction. The multilayer nature of the studied Au/Si films was also confirmed by X-ray reflectivity measurements, and with high-resolution transmission microscopy (HR-TEM) imaging on ultrathin Au/Si multilayers grown on a \textit{c}-Si substrate (see Supplemental Material, Fig. S1 \cite{Supplemental}).

To further investigate the dependence of spin Hall properties on the direction of the spin currents, we also used the ST-FMR technique which probes the spin currents that flow in the out-of-plane direction, as illustrated in Fig. 1(b). In our ST-FMR experiments, we investigated both single Au films and ultrathin Au/Si multilayers: three single polycrystalline Au (\textit{t}\textunderscript{Au}) films, where \textit{t}\textunderscript{Au} = 2, 3, 5 nm; and nine Au (\textit{t}\textunderscript{Au})/[Si (\textit{t}\textunderscript{Si})/Au (\textit{t}\textunderscript{Au})]\textunderscript{x4} multilayers, where \textit{t}\textunderscript{Au} = 2, 3, 4 nm and the thickness of the insulating Si interlayers \textit{t}\textunderscript{Si} = 2, 3, 4 nm. Each of these single Au films and Au/Si multilayers was sputter-deposited on a glass/Si (5 nm) substrate, and was capped with Ni\textunderscript{81}Fe\textunderscript{19} (4 nm)/SiN\textunderscript{x} (3 nm). The relatively low thickness of the Ni\textunderscript{81}Fe\textunderscript{19} layer is chosen to ensure that the radio-frequency (RF) charge current flows evenly through the Ni\textunderscript{81}Fe\textunderscript{19} and individual Au layers, while the 3-nm-thick SiN\textunderscript{x} capping layer is used to prevent oxidation of the Ni\textunderscript{81}Fe\textunderscript{19} layer (see Supplemental Material, Methods \cite{Supplemental}). 
\\

\begin{figure}[h]
\begin{center}
\scalebox{1}{\includegraphics[width=7.5 cm, clip]{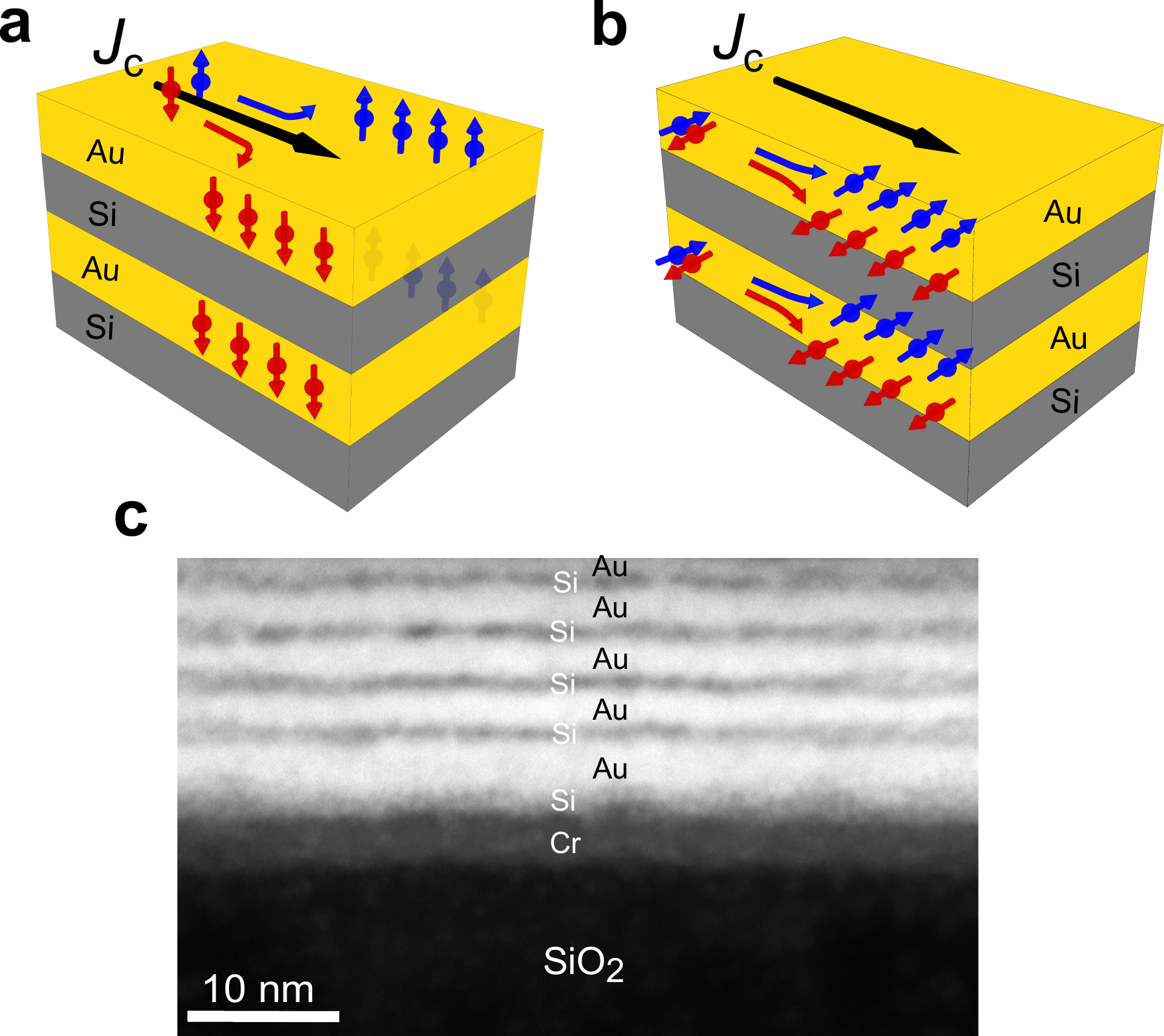}}
\end{center}
\caption{\label{sample} a) and b) Schematic representation of the spin Hall effect in ultrathin Au/Si multilayers showing the in- and out-of-plane spin currents, respectively. c) Dark-field cross-sectional STEM image of [Si (1.8 nm)/Au (2 nm)]\textunderscript{x5} multilayer stack sputtered on a glass/Cr (3 nm), showing well-defined layered structure. The lighter regions correspond to the Au layers, while the dark regions correspond to the Si interlayers and the Cr buffer layer.}
\end{figure}

\section{Results}
\subsection{Non-local transport measurements}
To probe the non-local transport, we patterned the four studied films into H-bar devices depicted in Fig. 2(a) using e-beam lithography and Ar ion-beam etching (see Supplemental Material, Methods \cite{Supplemental}). A top-view scanning electron microscope (SEM) image of the central region of such a H-bar device is shown in Fig. 2(b). The Hall bar device consists of six vertical wires of a width \textit{w} separated by a center-to-center distance \textit{L} and bridged by a horizontal wire of the same \textit{w}. For the two Au/Si multilayers and the single Au (10 nm) samples, \textit{w} is chosen to be 90 nm; this dimensionality is confirmed from SEM images, from which \textit{w} is measured to be (90 $\pm$ 5) nm. On the other hand, the width \textit{w} for the single Au (60 nm) sample is (110 $\pm$ 5) nm to be consistent with the previous study by Mihajlovi\'{c} \textit{et al.} on the non-local transport in Au (60 nm) H-bar structures \cite{Mihajlovic2009}. Moreover, the distance between the vertical wires \textit{L} is varied from (180 $\pm$ 10) nm to (550 $\pm$ 10) nm for all studied samples. To perform the non-local transport measurements, an unpolarized AC current is injected along the left vertical wire (\textit{y}-axis) while a non-local voltage is measured in the adjacent wire (\textit{y} direction). Three different transport mechanisms contribute to the non-local signal – the spin diffusive, the charge diffusive and the quasi-ballistic electron transport, as illustrated in Fig. 2(c), 2(d) and 2(e), respectively.

\begin{figure}[h]
\begin{center}
\scalebox{1}{\includegraphics[width=8 cm, clip]{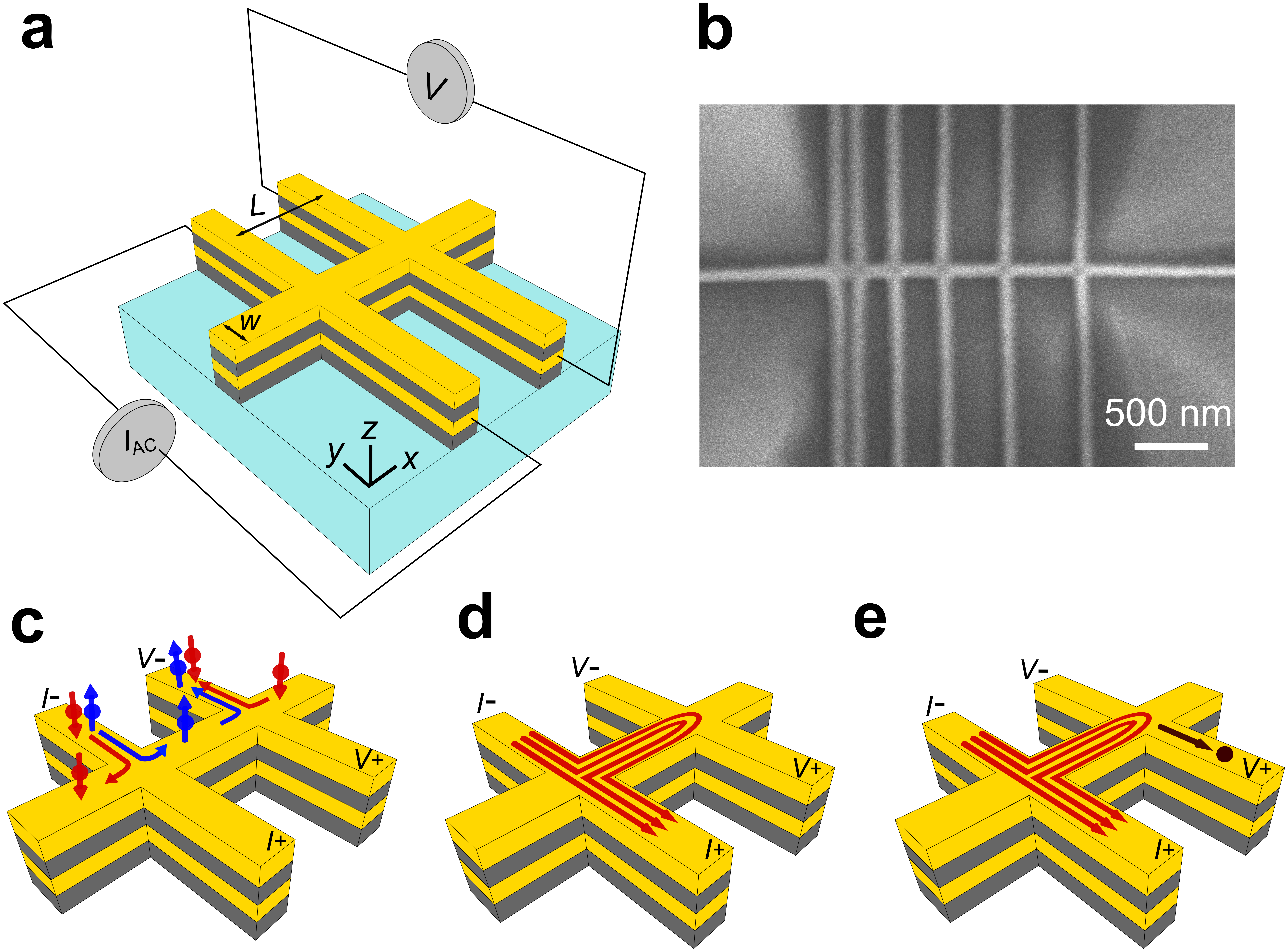}}
\end{center}
\caption{\label{sample} a) Experimental schematic of the non-local resistance measurement on Au/Si multilayers. b) Top-view SEM image of the Si (1.8 nm)/Au (2 nm) multilayer film patterned into a H-bar device with six vertical 90-nm-wide wires separated by various distances and bridged by a horizontal one. c), d) and e) Schematic representation of the three mechanisms involved in the non-local transport in H-bar structures, namely the spin diffusive, the charge diffusive and the quasi-ballistic transport, respectively.}
\end{figure}

In the previous experimental studies utilizing non-local transport, the spin diffusive contribution to the non-local signal was interpreted solely in terms of the bulk SHE \cite{Mihajlovic2009,Tian2016,Chen2019}. As shown in Fig. 2(c), the unpolarized charge current injected along the vertical wire induces a spin current in the horizontal wire due to the direct SHE. The spin current then diffuses and induces charge accumulation across the adjacent vertical wire via the inverse spin Hall effect (ISHE). Nevertheless, it was reported by Abanin \textit{et al.} \cite{Abanin2009} that the spin diffusive contribution can result not only from the bulk SHE, but also from interfacial SOC-related effects such as the Edelstein effect and the Rashba SOC. Unlike for bulk SHE, the charge current in systems with strong 2D Rashba SOC induces a spin accumulation rather than a spin current, which creates a charge current in the adjacent wire via the inverse Edelstein effect. In our study, the spin diffusive contribution to the non-local transport in single Au layers and ultrathin Au/Si multilayers will be interpreted as originating from both the bulk SHE and interfacial SOC-related effects. Therefore, the spin-to-charge conversion in the studied H-bar structures is characterized by a spin-charge conversion efficiency \textbf{$\theta$}\textunderscript{sc}, rather than by a bulk SHA \textbf{$\theta$}\textunderscript{SHE}. According to the model by Abanin \textit{et al.} \cite{Abanin2009}, the spin diffusive contribution to the non-local resistance for \textit{l}\textunderscript{e} $\leq$ \textit{w}, where \textit{l}\textunderscript{e} is electron mean free path, is expressed by:

\begin{equation}
\label{delta_H}
\ R_\mathrm{nl}^{sd}= \frac{1}{2} R_\mathrm{sq} \theta_\mathrm{sc}^{2}\frac{w}{l_\mathrm{s}} \exp \left( -\frac{L}{l_\mathrm{s}} \right)  
\end{equation}

where \textit{l}\textunderscript{s} is the spin diffusion length, \textbf{$\theta$}\textunderscript{sc} is the spin-charge conversion efficiency, and \textit{R}\textunderscript{sq} is the sheet resistance of the wire. In the case of [Au (\textit{t}\textunderscript{Au})/Si (\textit{t}\textunderscript{Si})]\textunderscript{N} multilayers where we assume that charge transport is negligible in Si, the sheet resistance equals $\frac{\rho_\mathrm{Au}}{N \textit{t}_\mathrm{Au}}$, where \textbf{$\rho$}\textunderscript{Au} is the resistivity of the individual Au layers, \textit{t}\textunderscript{Au} is the thickness of the individual Au layers, and \textit{N} is the number of repeats. 

In addition to the spin diffusive contribution, charge diffusion also contributes to the non-local signal when \textit{l}\textunderscript{e} $\leq$ \textit{w}. This corresponds to the spreading of the charge current density into the horizontal wire, leading to a non-zero voltage in the adjacent wire, as illustrated in Fig. 2(d). The charge diffusion contribution is defined as:

\begin{equation}
\label{delta_H}
\ R_\mathrm{nl}^{cd}= R_\mathrm{sq} \exp \left( -\pi \frac{L}{w} \right)  
\end{equation}

On the other hand, the electrons can also travel ballistically over the horizontal wire and then scatter ballistically into the lower region of the adjacent wire, thus generating a negative voltage as shown in Fig. 2(e). As described in \cite{Mihajlovic2009}, the quasi-ballistic contribution to the non-local signal can be expressed by:

\begin{equation}
\label{delta_H}
\ R_\mathrm{nl}^{b}= -b R_\mathrm{sq} \exp \left( - \frac{w}{l_\mathrm{e}} \right)\exp \left( -\pi \frac{L}{w} \right)  
\end{equation}

\begin{figure}[h]
\begin{center}
\scalebox{1}{\includegraphics[width=7 cm, clip]{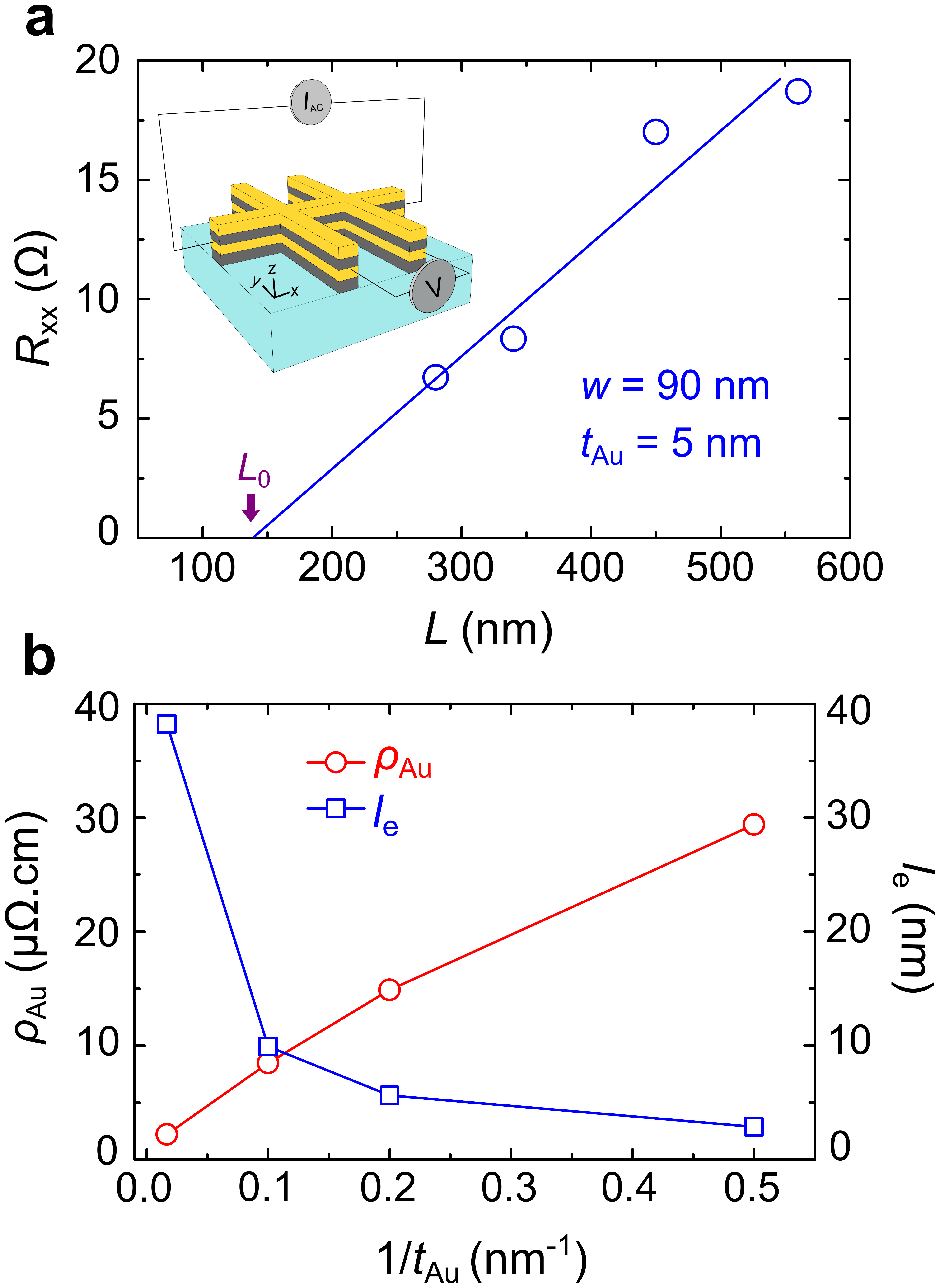}}
\end{center}
\caption{\label{sample} a) Resistance of each bridge wire between adjacent vertical wires of Au (5 nm)/Si (4 nm) multilayers as a function of \textit{L} at \textit{T} = 5 K. The blue solid line is a linear fit to the data. Upper inset: Experimental schematic of the local resistance measurement on Au/Si multilayers. b) Resistivity \textbf{$\rho$}\textunderscript{Au} and the corresponding electron mean free path \textit{l}\textunderscript{e} as a function of the inverse Au thickness 1/\textit{t}\textunderscript{Au} at 5 K, measured for the Au (\textit{t}\textunderscript{Au} = 10, 60 nm) films, as well as Au (\textit{t}\textunderscript{Au} = 2, 5 nm)/Si multilayers. The lines are guide to the eyes.}
\end{figure}

where \textit{l}\textunderscript{e} is the electron mean-free path and \textit{b} is a fitting parameter. Hence, measuring a negative non-local resistance in the studied H-bar structures would be a signature of the quasi-ballistic transport mechanism \cite{Mihajlovic2009}. To determine the sheet resistance \textit{R}\textunderscript{sq} of our four samples, we measured the local resistance \textit{R}\textunderscript{xx} as a function of temperature \textit{T} for each segment of the horizontal wire between the adjacent vertical wires, as shown in the inset of Fig. 3(a). One can see from Fig. 3(a) that the local resistance \textit{R}\textunderscript{xx} increases linearly with the distance \textit{L} at \textit{T} = 5 K. Moreover, the linear fit of \textit{R}\textunderscript{xx} crosses the \textit{L}-axis in a finite distance \textit{L}\textunderscript{0}, indicating that the effective distance between each of the adjacent vertical wires corresponds to \textit{L}\textunderscript{eff} = \textit{L} - \textit{L}\textunderscript{0}. This finding is in agreement with previous studies \cite{Mihajlovic2009}, and is attributed to the spreading of the charge current density into the vertical wires due to their finite width \cite{Mihajlovic2009}. For wires with \textit{w} = 90 nm, we measured an \textit{L}\textunderscript{0} of (135 $\pm$ 5) nm. In the following, we will use \textit{L} = \textit{L}\textunderscript{eff} in Equation 1, 2 and 3. 

To accurately estimate the resistivity of the Au layers \textbf{$\rho$}\textunderscript{Au}, we first measured the resistivity of the 3-nm-thick Cr buffer layer \textbf{$\rho$}\textunderscript{Cr} by performing temperature dependent resistance measurements on a glass/Cr (3 nm)/AlO\textunderscript{x} (3 nm) sample. The AlO\textunderscript{x} capping layer is used to prevent oxidation of the Cr layer. We found that the resistivity of the 3-nm-thick Cr layer decreases with increasing temperature from 250.1 \textbf{$\mu \Omega$}.cm at 5 K to 232.9 \textbf{$\mu \Omega$}.cm at 300 K. Therefore, the high resistivity of the 3-nm-thick Cr layer relative to that of Au indicates that most of the charge current flows through the Au layers, thus ruling out any strong contribution of the Cr layer to the non-local transport in the four investigated samples. Having considered the resistivity of the Cr buffer layer, we measured the temperature dependence of the resistivity of the individual Au layers (\textbf{$\rho$}\textunderscript{Au}) for the four studied samples. Fig. 3(b) shows that the resistivity \textbf{$\rho$}\textunderscript{Au} is inversely proportional to the thickness of the individual Au layers \textit{t}\textunderscript{Au}. Indeed, \textbf{$\rho$}\textunderscript{Au} strongly increases from a bulk value of 2.2 \textbf{$\mu \Omega$}.cm (resp. 3.79 \textbf{$\mu \Omega$}.cm) at 5 K (resp. 300 K) for the single Au (\textit{t}\textunderscript{Au} = 60 nm) sample, to a much higher value of 29.4 \textbf{$\mu \Omega$}.cm (resp. 35.27 \textbf{$\mu \Omega$}.cm) at 5 K (resp. 300 K) for Au (\textit{t}\textunderscript{Au} = 2 nm)/Si (1.8 nm) multilayers. Such an increase in resistivity for ultrathin Au is in agreement with previous studies \cite{Brangham2016}, and is mainly attributed to the dominance of surface scattering for ultrathin films. We further calculated the electron mean free path \textit{l}\textunderscript{e} values for the four studied samples using the Drude formula with an electron density for Au of $ n = 5.9 \times 10^{28}$ m\textsuperscript{-3} \cite{Mihajlovic2009}. We found that \textit{l}\textunderscript{e} strongly decreases from 38.2 nm (resp. 22.12 nm) at 5 K (resp. 300 K) for the single Au (60 nm) layer sample (in agreement with Ref. \cite{Mihajlovic2009}) to 2.95 nm (resp. 2.48 nm) at 5 K (resp. 300 K) for Au (\textit{t}\textunderscript{Au} = 2 nm)/Si (1.8 nm) multilayers. These findings are attributed to the strong increase of \textbf{$\rho$}\textunderscript{Au} in the ultrathin Au layers, and suggest that the quasi-ballistic contribution to the non-local resistance will be strongly suppressed in ultrathin Au-based multilayers. 

\begin{figure}[h]
\begin{center}
\scalebox{1}{\includegraphics[width=7 cm, clip]{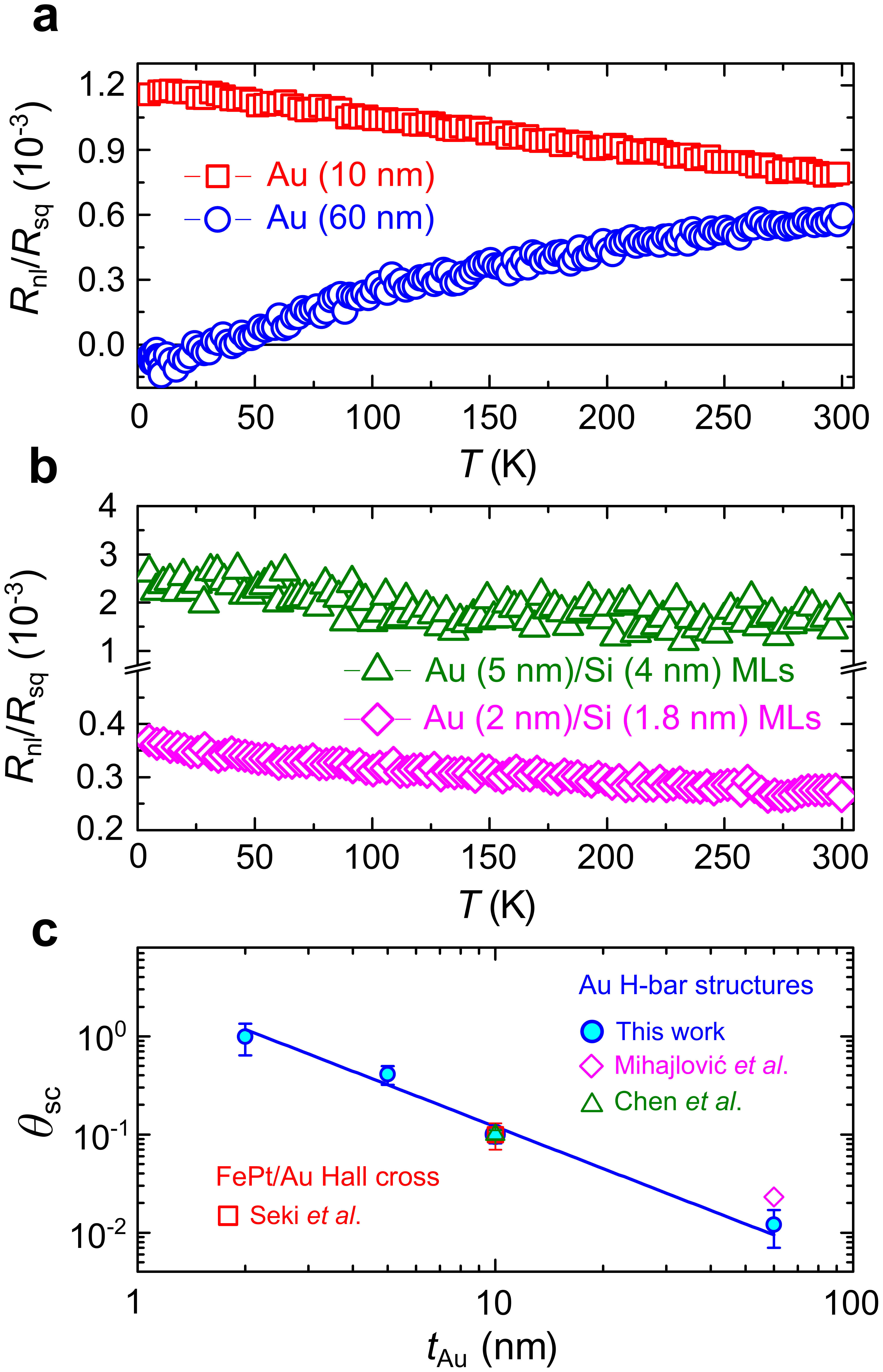}}
\end{center}
\caption{\label{sample} a) Temperature dependence of \textit{R}\textunderscript{nl}/\textit{R}\textunderscript{sq} measured for Au (\textit{t}\textunderscript{Au} = 10, 60 nm) films and for adjacent wires separated by a distance \textit{L} = 340 nm. b) Temperature dependence of \textit{R}\textunderscript{nl}/\textit{R}\textunderscript{sq} measured for Au (\textit{t}\textunderscript{Au} = 5 nm)/Si (4 nm) and Au (\textit{t}\textunderscript{Au} = 2 nm)/Si (1.8 nm) multilayers and for wires separated by a distance \textit{L} = 340 nm and 450 nm, respectively. c) Logarithmic plot of the spin-charge conversion efficiency \textbf{$\theta$}\textunderscript{sc} as a function of \textit{t}\textunderscript{Au} measured at \textit{T} = 5 K for the single Au (\textit{t}\textunderscript{Au}= 10, 60 nm) films, as well as Au (\textit{t}\textunderscript{Au} = 5 nm)/Si (4 nm) and Au (\textit{t}\textunderscript{Au} = 2 nm)/Si (1.8 nm) multilayers. The blue solid line is a fit to the non-local transport data. For comparison, we also plot the \textbf{$\theta$}\textunderscript{SHE} values of Au from literature. Red square: study of perpendicularly spin-polarized FePt/Au (10 nm) structure from Seki \textit{et al.} \cite{Seki2008}. Magenta diamond: non-local transport in Au (60 nm) H-bar structure from Mihajlovi\'{c} \textit{et al.} \cite{Mihajlovic2009}. Green triangle: non-local transport in Au (10 nm) H-bar structure from Chen \textit{et al.} \cite{Chen2019}.}
\end{figure}

We further measured the temperature dependence of the non-local resistance \textit{R}\textunderscript{nl} and extracted \textbf{$\theta$}\textunderscript{sc} for the four studied samples. Since all three different contributions to \textit{R}\textunderscript{nl} are proportional to the sheet resistance \textit{R}\textunderscript{sq}, we used the reduced non-local resistance \textit{R}\textunderscript{nl}/\textit{R}\textunderscript{sq} to investigate their temperature dependence. Fig. 4(a) displays the \textit{T} dependence of \textit{R}\textunderscript{nl}/\textit{R}\textunderscript{sq} measured for the single Au (60 nm) layer with \textit{L} = 340 nm. One can see from Fig. 4(a) that \textit{R}\textunderscript{nl}/\textit{R}\textunderscript{sq} decreases when \textit{T} is lowered and becomes negative around 33 K. This finding indicates that the non-local transport becomes dominated by the quasi-ballistic contribution at low temperatures in agreement with previous studies \cite{Mihajlovic2009,Chen2019}. In order to extract \textbf{$\theta$}\textunderscript{sc} of the 60-nm-thick Au layer, we followed the method of Mihajlovi\'{c} \textit{et al.} \cite{Mihajlovic2009} by plotting \textit{R}\textunderscript{nl}/\textit{R}\textunderscript{sq} as a function of \textit{l}\textunderscript{e}. As expected, we found that \textit{R}\textunderscript{nl}/\textit{R}\textunderscript{sq} fits with the quasi-ballistic and charge diffusive contributions to the non-local signal ($R_\mathrm{nl}^{cd}$ + $R_\mathrm{nl}^{b}$)/\textit{R}\textunderscript{sq}, where the fitting parameter \textit{b} = 22.14 is used (see Supplemental Material, Fig. S2). Thus, we deduced the \textit{T} dependence of $R_\mathrm{nl}^{sd}$/\textit{R}\textunderscript{sq} by subtracting ($R_\mathrm{nl}^{cd}$ + $R_\mathrm{nl}^{b}$)/\textit{R}\textunderscript{sq} from \textit{R}\textunderscript{nl}/\textit{R}\textunderscript{sq}. By assuming the spin diffusion length of Au \textit{l}\textunderscript{s} = 168 nm (resp. 65 nm) reported in Ref. \cite{Ku2006} (resp. Ref. \cite{Ji2004}), we extracted an upper limit for \textbf{$\theta$}\textunderscript{sc} of the Au (60 nm) sample of 0.012 (resp. 0.022) at 5 K, thus confirming the absence of a giant SHE in the studied Au (60 nm) sample. These \textbf{$\theta$}\textunderscript{sc} values are consistent with the work by Mihajlovi\'{c} \textit{et al.} \cite{Mihajlovic2009} and, more importantly, confirm the absence of any contribution from the 3-nm-thick Cr buffer layer to the non-local transport. \\

In contrast to the single Au (60 nm) layer, \textit{R}\textunderscript{nl}/\textit{R}\textunderscript{sq} measured for the single Au (10 nm) layer with \textit{L} = 340 nm is always positive and increases when \textit{T} is lowered, as shown in Fig. 4(a). This finding is attributed to the increase of the resistivity \textbf{$\rho$}\textunderscript{Au} in the Au (10 nm) layer, as shown in Fig. 3(b), leading to much lower \textit{l}\textunderscript{e} values ranging from 7.47 nm at 300 K to 9.92 nm at 5 K. Therefore, the contribution of ($R_\mathrm{nl}^{cd}$ + $R_\mathrm{nl}^{b}$)/\textit{R}\textunderscript{sq} to \textit{R}\textunderscript{nl}/\textit{R}\textunderscript{sq} measured for the Au (10 nm) layer is not as dominant as for the Au (60 nm) layer, which explains the difference observed in the \textit{T} dependence of \textit{R}\textunderscript{nl}/\textit{R}\textunderscript{sq} between these two samples. By assuming the b value of $\sim$ 22.14 , we found that ($R_\mathrm{nl}^{cd}$ + $R_\mathrm{nl}^{b}$)/\textit{R}\textunderscript{sq} corresponds to $7.18 \times 10^{-4}$ at 5 K for the Au (10 nm) layer. By subtracting ($R_\mathrm{nl}^{cd}$ + $R_\mathrm{nl}^{b}$)/\textit{R}\textunderscript{sq} from  \textit{R}\textunderscript{nl}/\textit{R}\textunderscript{sq}, we deduced a spin diffusion contribution $R_\mathrm{nl}^{sd}$/\textit{R}\textunderscript{sq} of $4.27 \times 10^{-4}$. To extract \textbf{$\theta$}\textunderscript{sc} of Au (10 nm), we assumed \textit{l}\textunderscript{s} = (75 $\pm$ 5) nm at 5 K, as demonstrated in a recent study on the non-local transport in Au (10 nm) \cite{Chen2019}. From this, we extracted \textbf{$\theta$}\textunderscript{sc} of 0.1 $\pm$ 0.05 for the Au (10 nm) layer, which is in good agreement with other previous studies \cite{Seki2008,Tian2016,Chen2019} as shown in Fig. 4(c). We now elucidate the non-local transport in the ultrathin Au/Si multilayers via the same approach previously used for the single Au layers. For Au (\textit{t}\textunderscript{Au} $\leq$ 5 nm) layers, the electron mean free path \textit{l}\textunderscript{e} is drastically reduced (\textit{l}\textunderscript{e} $\leq$ 6 nm) as previously discussed in Fig. 3(b). Hence, the quasi-ballistic contribution $R_\mathrm{nl}^{b}$/\textit{R}\textunderscript{sq} will be strongly suppressed for the ultrathin Au/Si multilayers. Fig. 4(b) shows that \textit{R}\textunderscript{nl}/\textit{R}\textunderscript{sq} measured for the Au (\textit{t}\textunderscript{Au} $\leq$ 5 nm)/Si multilayers is always positive and increases when \textit{T} is lowered, thus indicating that the non-local transport in these two multilayers is mostly dominated by the spin diffusion. \\

To extract \textbf{$\theta$}\textunderscript{sc} in these multilayers, we should first take into consideration the change of the spin diffusion length with the increase of \textbf{$\rho$}\textunderscript{Au} in the ultrathin Au. Indeed, it was previously reported in literature on materials with large \textit{l}\textunderscript{s} and spin relaxation via the Elliott-Yafet mechanism such as Au \cite{Ku2006} and Cu \cite{Villamor2013}, that \textit{l}\textunderscript{s} at low \textit{T} decreases with the resistivity following the opposite trend of the thickness \cite{Ku2006,Villamor2013}. On the other hand, we measured a strong increase of \textbf{$\rho$}\textunderscript{Au} at 5 K from the bulk value of 2.2 to 29.4 \textbf{$\mu \Omega$}.cm for 2-nm-thick Au. \textit{l}\textunderscript{s} is thus expected to be shorter for the ultrathin Au/Si multilayers. Therefore, it is reasonable to assume, for our estimation of a lower limit of \textbf{$\theta$}\textunderscript{sc}, that \textit{l}\textunderscript{s} corresponds to at most (43 $\pm$ 5) nm and (35 $\pm$ 5) nm at 5 K for each of the individual Au (5 nm) and Au (2 nm) layers, respectively. At \textit{T} = 5 K, we extracted the following values: \textbf{$\theta$}\textunderscript{sc} (\textit{t}\textunderscript{Au} = 5 nm) = 0.41 $\pm$ 0.09 and \textbf{$\theta$}\textunderscript{sc} (\textit{t}\textunderscript{Au} = 2 nm) = 0.99 $\pm$ 0.34, which are significantly larger than those extracted for the single Au (\textit{t}\textunderscript{Au} = 10, 60 nm) films. Furthermore, we extracted a room-temperature (RT) \textbf{$\theta$}\textunderscript{sc} value of 0.87 $\pm$ 0.34 for Au (2 nm)/Si (1.8 nm) multilayers by assuming \textit{l}\textunderscript{s} = (35 $\pm$ 5) nm, which is exceedingly larger than bulk \textbf{$\theta$}\textunderscript{SHE} values reported for other heavy metals, such as -0.33 for \textbf{$\beta$}-W (from \cite{Pai2012}), -0.12 for \textbf{$\beta$}-Ta (from \cite{Liu2012a}), and 0.1 for Pt (from \cite{Althammer2013,Zhang2013}). This implies that the extracted spin-charge conversion efficiency \textbf{$\theta$}\textunderscript{sc} from the spin diffusive term of the non-local resistance cannot be interpreted solely in terms of the bulk SHE. By plotting the \textbf{$\theta$}\textunderscript{sc} values measured for the four samples in a single graph on a logarithmic scale, one can see from Fig. 4(c) that \textbf{$\theta$}\textunderscript{sc} at \textit{T} = 5 K is strongly enhanced for ultrathin Au thickness. This finding implies that the thickness and, therefore, the resistivity have a decisive effect on the spin-to-charge conversion in Au. More importantly, this strongly suggests the coexistence of a bulk SHE ($ \theta_\mathrm{SHE}^{bulk} \approx \theta_\mathrm{sc}  (t_\mathrm{Au} = 60$ nm) = 0.012 $\pm$ 0.005) with a strong interfacial SOC effect which becomes dominant in ultrathin Au. 

\subsection{ST-FMR measurements} 

To give more insight into the microscopic origin of the SHE in the ultrathin Au layers, it is important to elucidate the anisotropy of the SHE with respect to the spin current flow direction. In this context, we further investigate the thickness-dependence of the SHE in Au by using the ST-FMR technique which probes the spin currents flow in the out-of-plane direction, as illustrated in Fig. 5(a). We studied both single Au (\textit{t}\textunderscript{Au})/Ni\textunderscript{81}Fe\textunderscript{19} (4 nm) bilayers, where \textit{t}\textunderscript{Au} = 2, 3, 5 nm; and ultrathin Au (\textit{t}\textunderscript{Au})/[Si (\textit{t}\textunderscript{Si})/Au (\textit{t}\textunderscript{Au})]\textunderscript{x4}/Ni\textunderscript{81}Fe\textunderscript{19} (4 nm) multilayers, where \textit{t}\textunderscript{Au} = 2, 3, 4 nm and \textit{t}\textunderscript{Si} = 2, 3, 4 nm. The magnetization \textit{M} of the 4-nm-thick Ni\textunderscript{81}Fe\textunderscript{19} layer is oriented in-plane. 

\begin{figure*}[!ht]
\center
\includegraphics[width=12 cm]{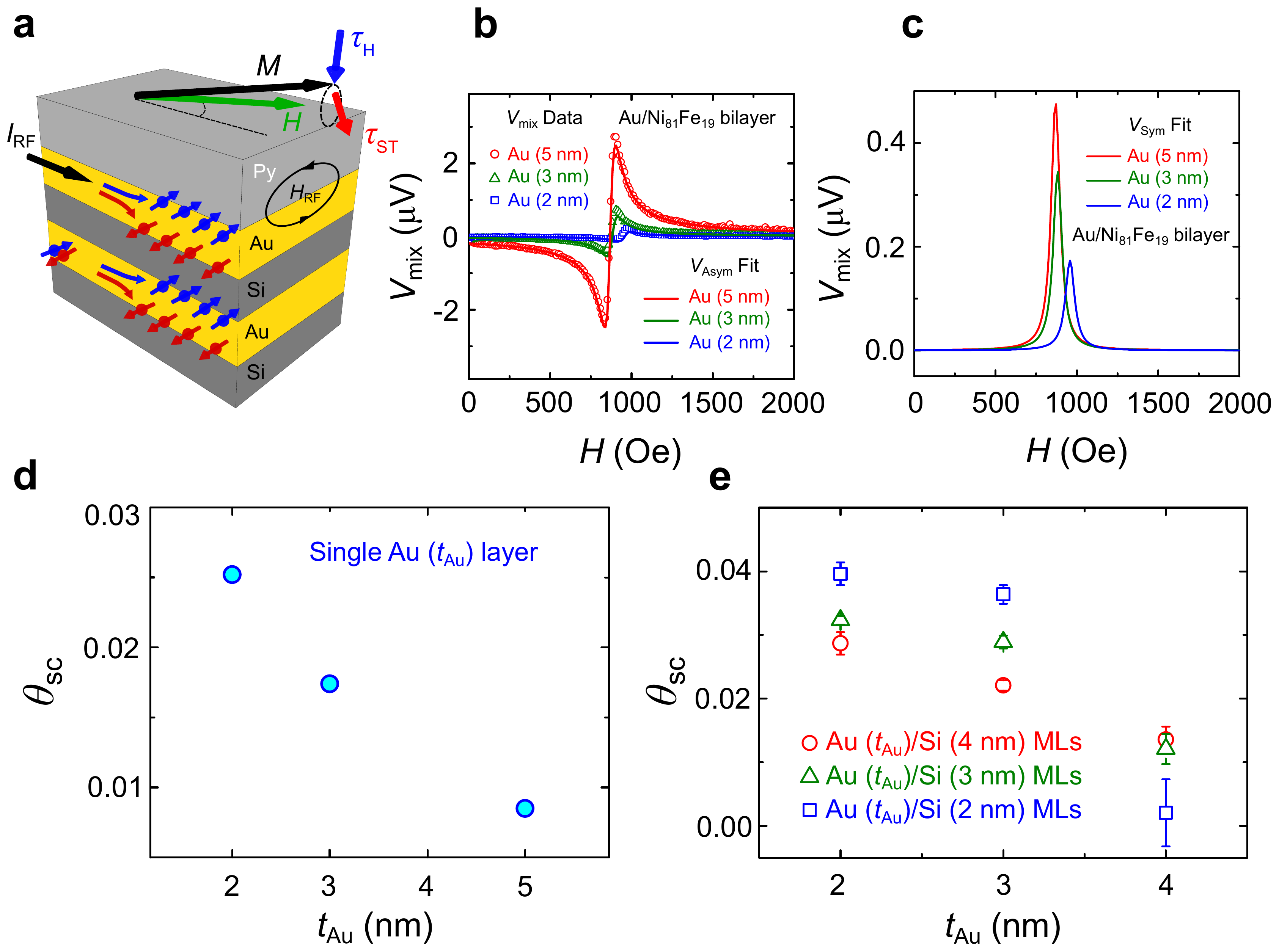}
\caption{\label{schema} ST-FMR measurement on single Au/Ni\textunderscript{81}Fe\textunderscript{19} bilayers and ultrathin [Si/Au]\textunderscript{N}/Ni\textunderscript{81}Fe\textunderscript{19} multilayer stacks. a) Schematic representation of the ST-FMR measurement in the ultrathin Au (\textit{t}\textunderscript{Au})/[Si (\textit{t}\textunderscript{Si})/Au (\textit{t}\textunderscript{Au})]\textunderscript{x4}/Ni\textunderscript{81}Fe\textunderscript{19} (4 nm) multilayer stack. \textit{I}\textunderscript{RF} and \textit{H}\textunderscript{RF} represent the applied radio-frequency (RF) current and the corresponding Oersted field, respectively. The out-of-plane spin current generated via the SHE in the top ultrathin Au layer induces a spin-transfer torque $\tau$\textunderscript{ST} in the Ni\textunderscript{81}Fe\textunderscript{19} layer. $\tau$\textunderscript{H} illustrates the torque induced by the Oersted field. \textit{H} and \textit{M} correspond to the external applied field and the magnetization of the Ni\textunderscript{81}Fe\textunderscript{19} layer, respectively. $45^{\circ}$ is the angle between applied field \textit{H} and the magnetization \textit{M}. b) and c) Representative ST-FMR resonance signals measured on Au (\textit{t}\textunderscript{Au})/Ni\textunderscript{81}Fe\textunderscript{19} (4 nm) under a driving frequency \textit{f} = 8 GHz and at room temperature, where \textit{t}\textunderscript{Au} = 2, 3 and 5 nm. The symbols in b) represent the measured DC voltage \textit{V}\textunderscript{mix} data. The solid lines in b) and c) represent the antisymmetric and symmetric Lorentzian fitting curves, respectively, whose sum fits the measured \textit{V}\textunderscript{mix} data. d) \textbf{$\theta$}\textunderscript{sc} measured for single Au (\textit{t}\textunderscript{Au})/Pt (4 nm) bilayers as a function of \textit{t}\textunderscript{Au}, where \textit{t}\textunderscript{Au} = 2, 3, 5 nm. e) \textbf{$\theta$}\textunderscript{sc} measured for ultrathin Au (\textit{t}\textunderscript{Au})/[Si (\textit{t}\textunderscript{Si})/Au (\textit{t}\textunderscript{Au})]\textunderscript{x4}/Ni\textunderscript{81}Fe\textunderscript{19} (4 nm) multilayers as a function of \textit{t}\textunderscript{Au}, where \textit{t}\textunderscript{Au} = 2, 3, 4 nm and \textit{t}\textunderscript{Si} = 2, 3, 4 nm.}
\end{figure*}

To probe the ST-FMR, we first patterned these studied films into 24-\textbf{$\mu$}m-wide and 88-\textbf{$\mu$}m-long microstrips using photolithography, and then fabricated symmetric waveguide contacts using DC sputtering and lift-off (see Supplemental Material, Methods \cite{Supplemental}). We applied an oscillating radio-frequency (RF) charge current \textit{I}\textunderscript{RF} at fixed frequencies (7-9 GHz) along the microstrips, and swept an external magnetic field \textit{H} in the in-plane direction at a $45^{\circ}$ angle with respect to the RF current direction, as illustrated in Fig. 5(a). An oscillating transverse spin current is then generated in each of the individual Au layers via the SHE. Hence, an oscillating spin-transfer torque induced by the out-of-plane spin current from the top Au layer is exerted on the Ni\textunderscript{81}Fe\textunderscript{19} layer, leading to magnetization precession and an oscillation of the anisotropic magnetoresistance of Ni\textunderscript{81}Fe\textunderscript{19}. This anisotropic magnetoresistance transforms the magnetization precession in Ni\textunderscript{81}Fe\textunderscript{19} into an RF resistance oscillation, which, by mixing with the RF current, generates a measurable dc voltage signal \textit{V}\textunderscript{mix} across the sample. By varying the strength of the applied magnetic field \textit{H}, the precession frequency of Ni\textunderscript{81}Fe\textunderscript{19} is controlled, creating a resonance in \textit{V}\textunderscript{mix}. To extract the spin-charge conversion efficiency of Au, we then use a lineshape analysis of the measured \textit{V}\textunderscript{mix}. Indeed, the resonance peak of the measured \textit{V}\textunderscript{mix} can be fitted by the sum of a symmetric Lorentzian \textit{F}\textunderscript{Sym} (\textit{H}) and an antisymmetric Lorentzian \textit{F}\textunderscript{Asym} (\textit{H}) as [\textit{S} \textit{F}\textunderscript{Sym} (\textit{H}) +  \textit{A} \textit{F}\textunderscript{Asym} (\textit{H})] \cite{Liu2011}. The symmetric component \textit{S} is a consequence of the spin Hall anti-damping torque induced by the out-of-plane spin current from the top Au layer. The antisymmetric component \textit{A} arises from the field-like torques, expected to be dominated by the Oersted field \textit{H}\textunderscript{RF} from the RF current in each of the individual Au layers. The ST-FMR resonance signal \textit{V}\textunderscript{mix} measured on single Au (\textit{t}\textunderscript{Au})/Ni\textunderscript{81}Fe\textunderscript{19} (4 nm) bilayers is exemplarily shown in Fig. 5(b). The corresponding antisymmetric and symmetric peaks are shown in Fig. 5(b) and Fig. 5(c), respectively.

We further quantitatively determined the \textbf{$\theta$}\textunderscript{sc} values in the studied samples by using the ratio of the symmetric component \textit{S} to the antisymmetric component \textit{A}. As described in Ref. \cite{Liu2011}, \textbf{$\theta$}\textunderscript{sc} of Au can be expressed by:

\begin{equation}
\label{delta_H}
\ \theta_\mathrm{sc} = \frac{S}{A}\frac{e \mu_\mathrm{0} M_\mathrm{s} t_\mathrm{NiFe} N t_\mathrm{Au}}{\hbar} \left(1 + \frac{4 \pi M_\mathrm{eff}}{H}\right)^\frac{1}{2}
\end{equation}

where \textbf{$\mu$}\textunderscript{0} is the permeability in vacuum, \textit{M}\textunderscript{s} is the saturation magnetization of Ni\textunderscript{81}Fe\textunderscript{19}, \textit{M}\textunderscript{eff} is the effective magnetization of Ni\textunderscript{81}Fe\textunderscript{19} which characterizes the out-of-plane demagnetization field, \textit{t}\textunderscript{NiFe} is the thickness of the Ni\textunderscript{81}Fe\textunderscript{19} layer, \textit{t}\textunderscript{Au} is the thickness of the individual Au layers, \textit{N} is the number of the individual Au layers, and \textit{H} is the external magnetic field. For each of the studied samples, \textit{M}\textunderscript{s} is measured with vibrating sample magnetometry (VSM), while $4 \pi M_\mathrm{eff}$ is extracted by fitting the frequency dependence of the resonance field to the Kittel equation \cite{Liu2011}. We found that the values \textit{M}\textunderscript{s} and \textit{M}\textunderscript{eff} are very similar, around $6.7 \times 10^{5}$ A/m, which is mainly due to the strong in-plane anisotropy of the Ni\textunderscript{81}Fe\textunderscript{19} layer.

Fig. 5(d) displays the Au thickness dependency of the room-temperature \textbf{$\theta$}\textunderscript{sc} values measured for single Au (\textit{t}\textunderscript{Au})/Ni\textunderscript{81}Fe\textunderscript{19} (4 nm) bilayers, where \textit{t}\textunderscript{Au} = 2, 3, 5 nm. One can see from Fig. 5(d) that \textbf{$\theta$}\textunderscript{sc} strongly increases from $8.5 \times 10^{-3}$ to 0.025 when \textit{t}\textunderscript{Au} is scaled down from 5 to 2 nm, a trend that is consistent with the non-local transport results. Nevertheless, these room-temperature \textbf{$\theta$}\textunderscript{sc} values obtained from the ST-FMR are much smaller than those extracted from the non-local transport at \textit{T} = 5 K. First, this finding can be attributed to the decrease of \textbf{$\theta$}\textunderscript{sc} with \textit{T}, as reported in a previous study on Au \cite{Isasa2015}. Second, the \textbf{$\theta$}\textunderscript{sc} values obtained from the ST-FMR do not account for the reduction of the spin transparency at the Au/Ni\textunderscript{81}Fe\textunderscript{19} interface originating from the spin backflow (SBF) and the spin memory loss (SML) \cite{Pai2015}, which is thought to often reduce of the out-of-plane spin torques by at least a factor of two. Hence, it is reasonable to attribute the discrepancy between the \textbf{$\theta$}\textunderscript{sc} values obtained with both ST-FMR and non-local transport techniques to the reduction of the spin transparency at the Au/Ni\textunderscript{81}Fe\textunderscript{19} interface. On the other hand, these findings might also suggest that the SHE in ultrathin Au is anisotropic, favoring more efficient spin-to-charge conversion in the in-plane direction rather than the out-of-plane direction. We further extracted the room-temperature \textbf{$\theta$}\textunderscript{sc} values in ultrathin Au (\textit{t}\textunderscript{Au})/[Si (\textit{t}\textunderscript{Si})/Au (\textit{t}\textunderscript{Au})]\textunderscript{x4}/Ni\textunderscript{81}Fe\textunderscript{19} (4 nm) multilayers, as shown in Fig. 5(e). Similarly to the single Au/Ni\textunderscript{81}Fe\textunderscript{19} bilayers, \textbf{$\theta$}\textunderscript{sc} increases from 0.012 to 0.039 by scaling down \textit{t}\textunderscript{Au} from 4 to 2 nm in ultrathin [Si/Au]\textunderscript{N}/Ni\textunderscript{81}Fe\textunderscript{19} multilayers. Note that these \textbf{$\theta$}\textunderscript{sc} values are slightly larger than those obtained for single Au/Ni\textunderscript{81}Fe\textunderscript{19} bilayers. This finding may be explained by the enhanced surface scattering in [Si/Au]\textunderscript{N}/Ni\textunderscript{81}Fe\textunderscript{19} multilayers due to the sample roughness, which enhances the extrinsic contribution to the SHE and, therefore, the \textbf{$\theta$}\textunderscript{sc} value. 

\section{Discussion}

The sharp increase of the \textbf{$\theta$}\textunderscript{sc} in ultrathin Au, demonstrated by both non-local transport and ST-FMR techniques, calls into question the interpretation of the spin-to-charge conversion as originating solely from the bulk SHE, and suggests the coexistence of both bulk and interfacial mechanisms. To further elucidate the microscopic origin of the spin-to-charge conversion in Au, we discuss the thickness dependence of \textbf{$\theta$}\textunderscript{sc} by fully accounting for both intrinsic and extrinsic SHEs. Owing to the intrinsic SOC in the band structure, the intrinsic SHA in Au can be expressed by \textbf{$\theta$}\textunderscript{SHE}\textsuperscript{intrinsic} = \textbf{$\sigma$}\textunderscript{SHE}\textsuperscript{intrinsic}/\textbf{$\sigma$}\textunderscript{Au}, where \textbf{$\sigma$}\textunderscript{SHE}\textsuperscript{intrinsic} is the intrinsic SHE conductivity and \textbf{$\sigma$}\textunderscript{Au} is the longitudinal conductivity of Au. Such an intrinsic contribution can be the dominant mechanism in a moderately dirty metal, as reported in an experimental study on Pt by Sagasta \textit{et al.} \cite{Sagasta2016}. Since scaling down the Au thickness to a few nm is accompanied by a strong increase of the resistivity \textbf{$\rho$}\textunderscript{Au}, one may intuitively attribute the strong enhancement of \textbf{$\theta$}\textunderscript{sc} in ultrathin Au to the intrinsic SHE mechanism. Nevertheless, the intrinsic SHE conductivity of Au is relatively small due to the small \textit{d}-electron density of states at the Fermi level, in contrast to that of Pt \cite{Tanaka2008}. Indeed, theoretical studies on Au have predicted relatively small \textbf{$\sigma$}\textunderscript{SHE}\textsuperscript{intrinsic} values, namely 400 $\Omega^{-1}$.cm$^{-1}$ \cite{Guo2009} and 731 $\Omega^{-1}$.cm$^{-1}$ \cite{Yao2005}. To verify the role of the intrinsic SHE in Au, we plot in Fig. 6 the measured \textbf{$\theta$}\textunderscript{sc} values and the theoretically predicted \textbf{$\sigma$}\textunderscript{SHE}\textsuperscript{intrinsic} as a function of \textbf{$\sigma$}\textunderscript{Au} on a logarithmic scale. Despite the discrepancy between the \textbf{$\theta$}\textunderscript{sc} values measured with non-local transport and ST-FMR techniques, one can see from Fig. 6 that logarithmic slopes of \textbf{$\theta$}\textunderscript{sc} for both techniques are very similar. Moreover, Fig. 6 shows that \textbf{$\theta$}\textunderscript{SHE}\textsuperscript{intrinsic} increases in ultrathin Au with low \textbf{$\sigma$}\textunderscript{Au} values, with a logarithmic slope smaller than the one of the measured \textbf{$\theta$}\textunderscript{sc}. More importantly, the predicted \textbf{$\theta$}\textunderscript{SHE}\textsuperscript{intrinsic} values in ultrathin Au are still negligible compared to the \textbf{$\theta$}\textunderscript{sc} values measured with the non-local transport as well as the SHA values reported in previous studies \cite{Seki2008,Mihajlovic2009,Chen2019}. Therefore, the intrinsic SHE contribution is negligible and cannot explain the strong increase of \textbf{$\theta$}\textunderscript{sc} in ultrathin Au.

\begin{figure}[h]
\begin{center}
\scalebox{1}{\includegraphics[width=8 cm, clip]{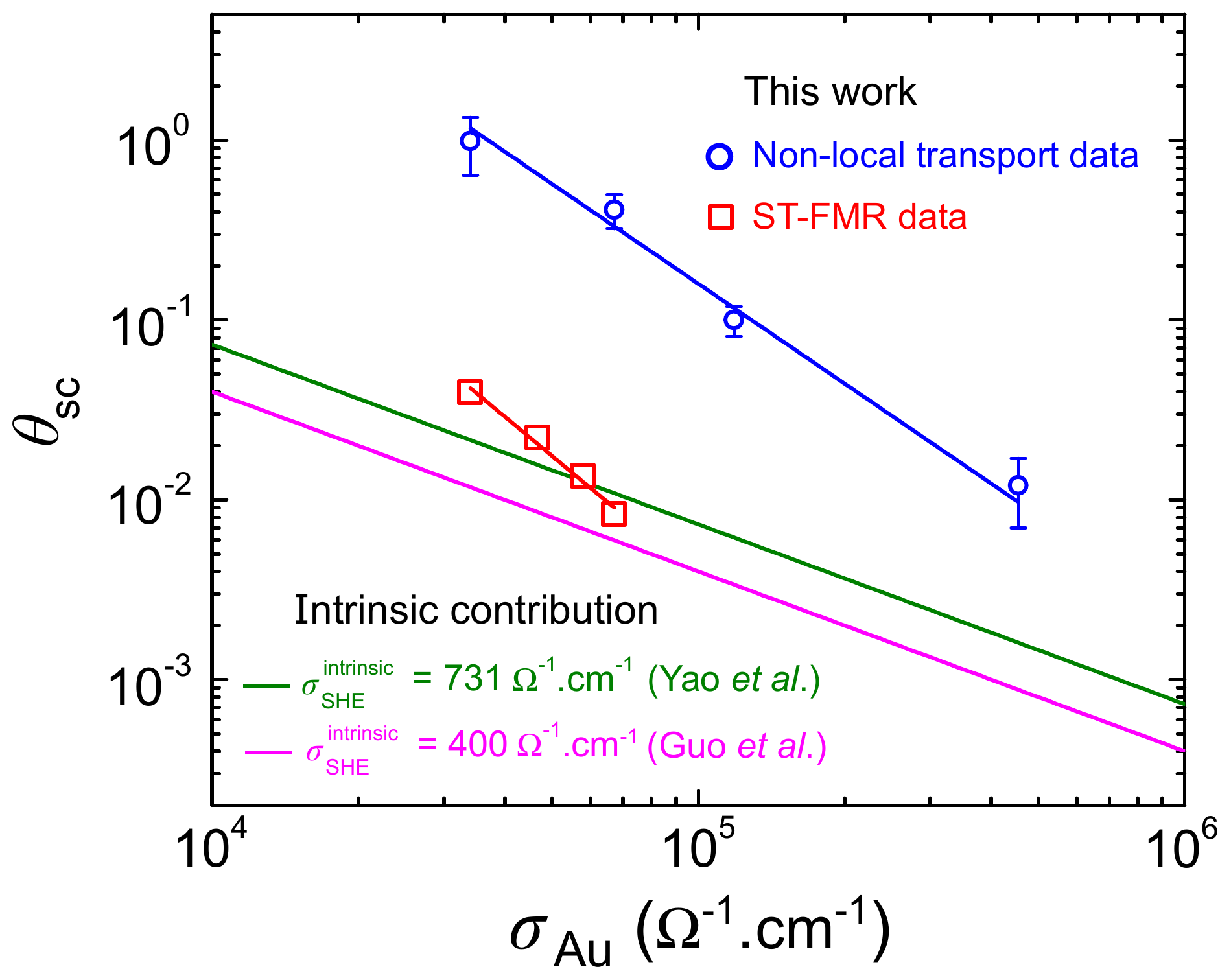}}
\end{center}
\caption{\label{sample} Logarithmic plot of the spin-charge conversion efficiency θsc as a function of the longitudinal conductivity \textbf{$\sigma$}\textunderscript{Au} of Au for the samples investigated in this work with non-local transport and ST-FMR measurements. The blue and red solid lines are a fit to the non-local transport and ST-FMR data, respectively. The green and magenta solid lines correspond to the intrinsic contribution to the SHA in Au \textbf{$\theta$}\textunderscript{SHE}\textsuperscript{intrinsic} = \textbf{$\sigma$}\textunderscript{SHE}\textsuperscript{intrinsic}/\textbf{$\sigma$}\textunderscript{Au} using \textbf{$\sigma$}\textunderscript{SHE}\textsuperscript{intrinsic} = 400 $\Omega^{-1}$.cm$^{-1}$ from Guo \textit{et al.} \cite{Guo2009} and \textbf{$\sigma$}\textunderscript{SHE}\textsuperscript{intrinsic} = 731 $\Omega^{-1}$.cm$^{-1}$ from Yao \textit{et al.} \cite{Yao2005}
}
\end{figure}

The strong increase of \textbf{$\theta$}\textunderscript{sc} is also unlikely to be explained by the extrinsic skew scattering SHE, since this extrinsic contribution is dominant only for clean metals \cite{Hoffmann2013} and should decrease in the relatively highly resistive ultrathin Au. Nevertheless, previous experimental and theoretical studies on the SHE in Au using perpendicularly spin-polarized FePt/Au structures revealed the key role of the extrinsic skew scattering contribution \cite{Seki2008,Seki2010,Gu2010,Guo2009}. Indeed, it was reported that the large SHA of $\sim$0.11 measured in FePt/Au (10 nm) structure originates from the extrinsic surface-assisted skew scattering due to Pt impurities \cite{Seki2008,Seki2010,Gu2010} and/or Fe Kondo impurities \cite{Guo2009}. However, we measured a similar \textbf{$\theta$}\textunderscript{sc} value in Au (10 nm) H-bar structures with the non-local transport, in agreement with Ref. \cite{Chen2019}, where both magnetic and strong SOC impurities are absent. This finding rules out the interpretation that the strong increase of \textbf{$\theta$}\textunderscript{sc} in ultrathin Au layers arises from the previously reported surface-assisted skew scattering mechanism \cite{Seki2008,Seki2010,Gu2010,Guo2009}. Furthermore, it was shown in a recent theoretical study that the surface scattering combined with a strong interfacial SOC can lead to a SHE that subscribes only to the extrinsic side-jump SHE, which is usually dominant for moderately dirty metals \cite{Zhou2015}. Hence, this extrinsic contribution could explain the thickness-dependence of \textbf{$\theta$}\textunderscript{sc} in Au; however, it is unlikely to explain the large \textbf{$\theta$}\textunderscript{sc} values measured in ultrathin Au with the non-local transport. 

Previous angle-resolved photoemission spectroscopy (ARPES) studies revealed a large Rashba-type splitting on Au(111) surface \cite{LaShell1996,Nicolay2001,Nechaev2009}, indicating the presence of a strong interfacial Rashba SOC. More importantly, it has been shown in a previous experimental study by Bondarenko \textit{et al.} that the Au/Si(111) interface can have metallic spin-split surface states with a large spin-splitting energy of 190 meV \cite{Bondarenko2013}. Since the sputter-deposited Au/Si multilayer has a strong (111) out-of-plane texture as shown in Ref. \cite{Kan2014}, a strong Rashba spin-orbit splitting should be present at the Au/Si interfaces in the studied Au/Si multilayers. Note that since the sputter-growth of Au on Si differs from the one of Si on Au, the Au/Si and Si/Au interfaces are never quite equivalent. Hence, the Rashba-Edelstein contributions from opposing interfaces in the investigated Au/Si multilayers would not cancel out. On the other hand, the spin-splitting energy at the Au/Si(111) interface is as large as the one reported for the Bi/Ag(111) interface, i.e., 200 meV \cite{Ast2007}. Moreover, Rojas-S\'{a}nchez \textit{et al.} experimentally demonstrated a large spin-to-charge conversion at the Bi/Ag Rashba interface \cite{Rojas-Sanchez2013}, which yields a SHA of 1.5 when interpreted only by the bulk SHE. Taking these findings into account, the increase of \textbf{$\theta$}\textunderscript{sc} in ultrathin Au to exceptionally large values ($\sim$0.99 for \textit{t}\textunderscript{Au} = 2 nm) can be plausibly explained by the Edelstein effect, arising from the strong interfacial Rashba SOC, and not from the bulk SHE. 

\section{Conclusion}

In conclusion, we have experimentally investigated the thickness dependence of the spin-charge conversion efficiency \textbf{$\theta$}\textunderscript{sc} in single Au films and ultrathin Au/Si multilayers with two different techniques, namely non-local transport and ST-FMR. We first found that the \textbf{$\theta$}\textunderscript{sc} values measured for the single Au (\textit{t}\textunderscript{Au} = 10, 60 nm) layers are consistent with the literature. Moreover, we demonstrated that \textbf{$\theta$}\textunderscript{sc} of Au measured with the non-local transport in ultrathin Au (\textit{t}\textunderscript{Au} = 2, 5 nm)/Si multilayers is strongly enhanced and reaches exceedingly large values. A similar thickness-dependent behavior of \textbf{$\theta$}\textunderscript{sc} was obtained using the ST-FMR technique, however, with much lower \textbf{$\theta$}\textunderscript{sc} values. Our experimental results evidence the coexistence of a strong interfacial SOC effect which becomes dominant in ultrathin Au, and bulk SHE with a relatively low bulk SHA. More importantly, these findings suggest the key role of the Rashba-Edelstein effect in the spin-to-charge conversion in ultrathin Au, and help pave the way for the use of ultrathin Au in emerging spintronic devices.

\noindent
\textbf{Acknowledgments} 

\noindent
The authors would like to thank A. Fert, G. Mihajlovi\'{c}, M. Stiles, V. P. Amin and F. Casanova for fruitful discussions, and R. Descoteaux from CMRR for technical assistance. The sample preparation was supported by the National Science Foundation under Grant No. NSF-DMR-1610538. The sample characterization, measurements, and data analysis were supported as part of Quantum Materials for Energy Efficient Neuromorphic Computing, an Energy Frontier Research Center funded by the U.S. DOE, Office of Science.\\

\noindent
\textbf{Additional information} 

\noindent
The authors declare no competing financial interests.

\renewcommand{\figurename}{\textbf{Supplemental Figure}}

\newpage
\maketitle

\begin{center}
\Large{\textbf{Supplemental Material}}
\vspace{1 cm}
\end{center}

\begin{center}
\begin{large}
\textbf{Methods}
\end{large}
\end{center}

\textbf{A. Thin film deposition} \\

Ultrathin Au/Si multilayers were grown by alternately DC magnetron sputtering Si and Au. Sputtering rates for Au at 50 W and Si at 100 W were 1.15 {\AA}/s and 0.54 {\AA}/s, respectively, as determined by low-angle X-ray reflectivity measurements of calibration sample film thicknesses. In our non-local transport experiments, four different samples were studied: two single polycrystalline Au films (Au (10 nm) and Au (60 nm)); and two Au/Si multilayers ([Si (1.8 nm)/Au (2 nm)]\textunderscript{x5} and Au (5 nm)/[Si (4 nm)/Au (5 nm)]\textunderscript{x5}). Each sample was deposited onto a glass/Cr (3 nm) substrate, where the 3-nm-thick buffer layer is used to ensure good adhesion of the single Au layers and the Au/Si multilayers to the substrate. In our ST-FMR experiments, both single Au films and ultrathin Au/Si multilayers were studied: three single polycrystalline Au (\textit{t}\textunderscript{Au}) films, where \textit{t}\textunderscript{Au} = 2, 3, 5 nm; and nine Au/Si multilayers with the following stacking structure: Au (\textit{t}\textunderscript{Au})/[Si (\textit{t}\textunderscript{Si})/Au (\textit{t}\textunderscript{Au})]\textunderscript{x4}, where \textit{t}\textunderscript{Au} = 2, 3, 4 nm and \textit{t}\textunderscript{Si} = 2, 3, 4 nm. Each of these samples was deposited on a glass/Si (5 nm) substrate and capped with Ni\textunderscript{81}Fe\textunderscript{19} (4 nm)/SiN\textunderscript{x} (3 nm). The SiN\textunderscript{x} layer helps prevent Ni\textunderscript{81}Fe\textunderscript{19} oxidation. The Ni\textunderscript{81}Fe\textunderscript{19} layers were grown by DC sputtering, while the SiN\textunderscript{x} layers grown by RF sputtering. Sputtering rates for Ni\textunderscript{81}Fe\textunderscript{19} at 100 W and SiN\textunderscript{x} at 150 W were 0.49 {\AA}/s and 0.092 {\AA}/s, respectively. The base pressure of the chamber was 5$\times$10\textsuperscript{-8} Torr, and all depositions were performed at room temperature in an Ar gas atmosphere with a pressure fixed at 2.7 mTorr. \\

\textbf{B. Device fabrication and characterization}\\

In our non-local transport experiments, the four studied films were patterned into H-bar devices using a PMMA (100 nm)/HSQ (90 nm) bilayer resist e-beam lithography process. The films were then etched to the glass top surface using Ar ion beam etching. Photolithography, DC sputtering and lift-off techniques were then used to fabricate the top contact pads consisting of Ti (10 nm)/Au (150 nm). Transport experiments on H-bar structures were conducted in a Physical Property Measurement System (PPMS) (Quantum Design, Inc.), where the temperature was varied from 5 to 300 K. A resistance bridge (Model 370 AC Resistance Bridge, LakeShore Cryotronics, Inc.) was used to measure both non-local and local resistances. \\

In our ST-FMR experiments, room-temperature magnetic hysteresis loops were first measured on continuous films by vibrating sample magnetometry (Quantum Design, Inc.). Slope-correction of the hysteresis loops was conducted to account for diamagnetic responses from the substrates by subtracting constant negative slopes at high fields from each dataset. The studied films were then patterned into 24-$\mu$m-wide and 88-$\mu$m-long microstrips using photolithography and Ar ion beam etching. Symmetric waveguide contacts consisting of Ti (10 nm)/Au (150 nm) were fabricated using photolithography, DC sputtering and lift-off techniques. An oscillating RF charge current was then applied at fixed frequencies (7-9 GHz) along the microstrips, and an external magnetic field up to 2000 Oe was applied in the in-plane direction at a $45^{\circ}$ angle with respect to the RF current direction. The DC voltage signal generated across the microstrips was then measured as a function of the applied magnetic field, and a lineshape analysis was used to extract the spin-charge conversion efficiency of Au.

\newpage
\begin{center}
\begin{large}
\textbf{High resolution transmission electron microscopy (HR-TEM) image of ultrathin Au/Si multilayers.}
\end{large}
\end{center}

\begin{figure}[h]
\begin{center}
\scalebox{1}{\includegraphics[width=11 cm, clip]{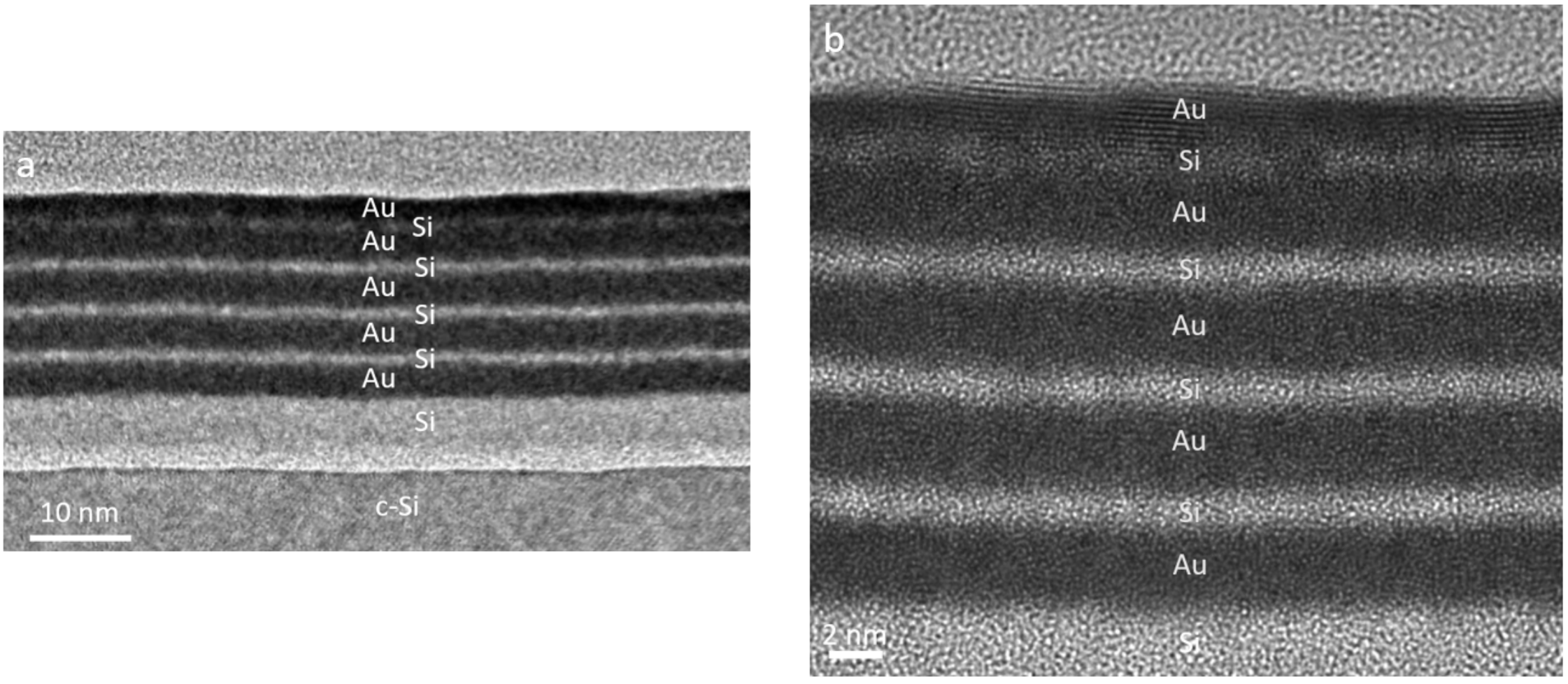}}
\end{center}
\makeatletter 
\renewcommand{\thefigure}{\textbf{1}}
\caption{\label{sample} The HR-TEM images were obtained on a FEI Talos F200X TEM/STEM, operating at 200 kV. An $\sim$ 30-nm-thick cross-section of an ultrathin Au/Si multilayer film grown on a \textit{c}-Si substrate with a 5-nm-thick amorphous Si buffer layer was fabricated via lift-out using a Zeiss NVision40 focused ion-beam (FIB) milling system. A protective carbon layer was used to prevent damage to the sample while being exposed to the Ga ion beam during FIB milling. The lighter regions correspond to Si layers, and the dark regions to Au layers. The HR-TEM images reveal that the top three layers are partially crystalline. The crystalline Au grains have a [111]-type direction along the growth direction. The lack of a sharp change in intensity between Au and Si layers is indicative of elemental intermixing at the interfaces.
}
\end{figure}

\newpage
\begin{center}
\begin{large}
\textbf{Fitting of the non-local resistance measured for 60-nm-thick Au film}
\end{large}
\end{center}

We followed the method of Mihajlovi\'{c} \textit{et al.} \cite{Mihajlovic2009} by plotting \textit{R}\textunderscript{nl}/\textit{R}\textunderscript{sq} as a function of \textit{l}\textunderscript{e} and fitting the experimental curve with the following equation:

\begin{equation}
\label{delta_H}
\ \frac{R_\mathrm{nl}}{R_\mathrm{sq}} = a \left(1 - b \exp \left(- \frac{w}{l_\mathrm{e}}\right) \right) 
\end{equation}

where $a = \exp \left( -\pi \frac{L - L_\mathrm{0}}{w} \right)$ with \textit{L} = 340 nm, \textit{L}\textunderscript{0} = 83 nm and \textit{w} = 110 nm, and the value of fitting parameter \textit{b} is 22.14 (see Fig. S2). This fit indicates that the measured \textit{R}\textunderscript{nl}/\textit{R}\textunderscript{sq} corresponds to the quasi-ballistic as well as the charge diffusive contributions to the non-local signal ($R_\mathrm{nl}^{cd}$ + $R_\mathrm{nl}^{b}$)/\textit{R}\textunderscript{sq}, confirming the absence of a giant SHE in the studied Au (60 nm) sample.

\begin{figure}[h]
\begin{center}
\scalebox{1}{\includegraphics[width=8 cm, clip]{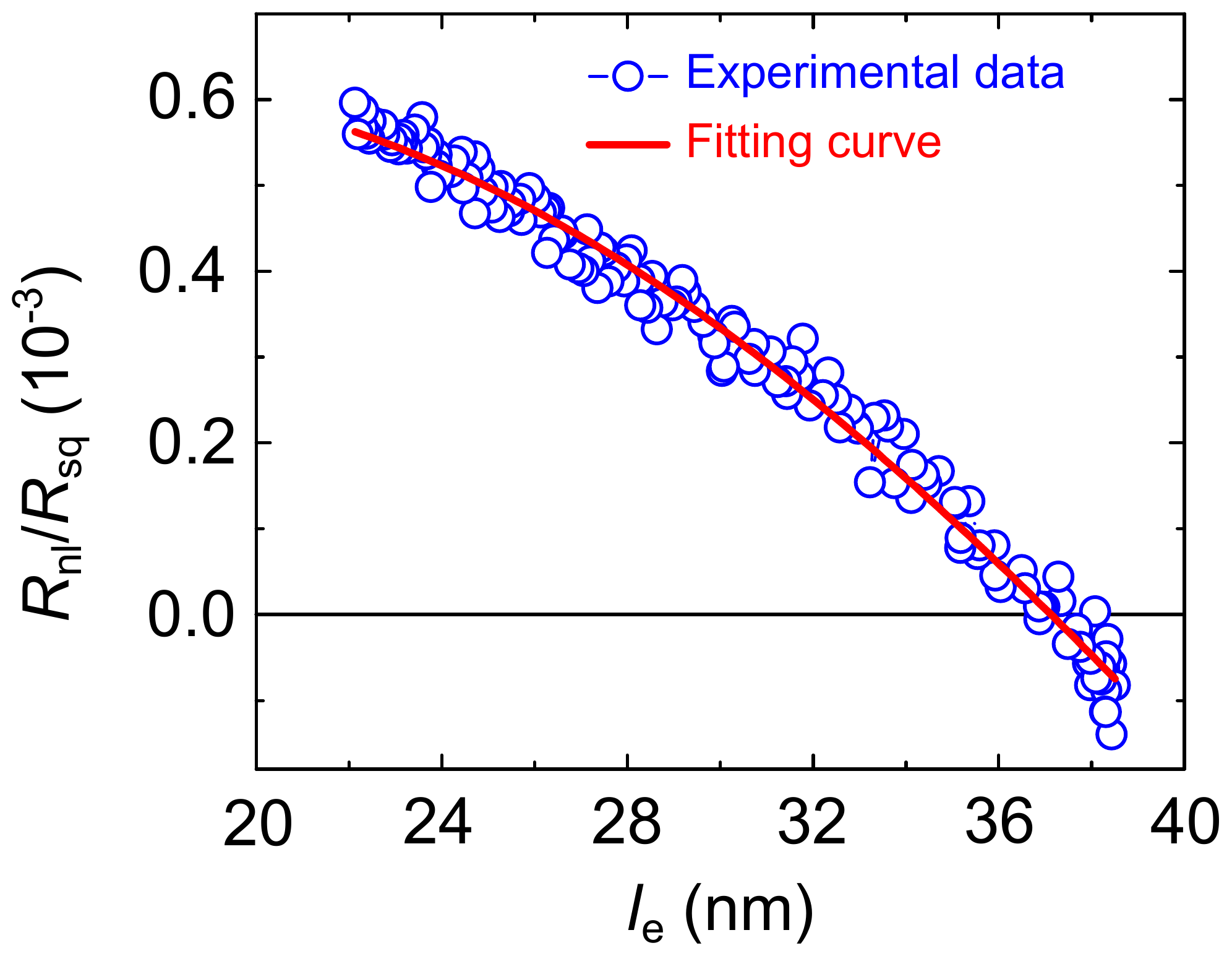}}
\end{center}
\makeatletter 
\renewcommand{\thefigure}{\textbf{2}}
\caption{\label{sample} \textit{R}\textunderscript{nl}/\textit{R}\textunderscript{sq} as a function of the electron mean free path measured for the studied Au (60 nm) film with adjacent wires separated by a distance \textit{L} = 340 nm. Corresponding fit to Eq. 1 is shown as red line.}
\end{figure}


\begin{thebibliography}{100}

% 1
\bibitem{Hellman2017}
F. Hellman, A. Hoffmann, Y. Tserkovnyak, G. S. D. Beach, E. E. Fullerton, C. Leighton, A. H. MacDonald, D. C. Ralph, D. A. Arena, H. A. D\"{u}rr, P. Fischer, J. Grollier, J. P. Heremans, T. Jungwirth, A. V. Kimel, B. Koopmans, I. N. Krivorotov, S. J. May, A. K. Petford-Long, J. M. Rondinelli, N. Samarth, I. K. Schuller, A. N. Slavin, M. D. Stiles, O. Tchernyshyov, A. Thiaville, B. L. Zink, \textit{Rev. Mod. Phys.} \textbf{89}, 025006 (2017).
% 2
\bibitem{Soumyanarayanan2016}
A. Soumyanarayanan, N. Reyren, A. Fert, C. Panagopoulos, \textit{Nature} \textbf{539}, 509-517 (2016).
% 3
\bibitem{Manchon2015}
A. Manchon, H. C. Koo, J. Nitta, S. M. Frolov, R. A. Duine, \textit{Nat. Mater.}  \textbf{14}, 871-882 (2015).
% 4
\bibitem{Sklenar2016}
J. Sklenar, W. Zhang, M. B. Jungfleisch, W. Wiang, H. Saglam, J. E. Pearson, J. B. Ketterson, A. Hoffmann, \textit{J. Appl. Phys.} \textbf{120}, 180901 (2016).
% 5
\bibitem{Hoffmann2013}
A. Hoffmann, \textit{IEEE Trans. Magn.} \textbf{49}, 10 (2013).
% 6
\bibitem{Sinova2015}
J. Sinova, S. O. Valenzuela, J. Wunderlich, C. H. Back, T. Jungwirth, \textit{Rev. Mod. Phys.} \textbf{87}, 1213 (2015).
% 7
\bibitem{Dyakonov1971}
M. I. D'yakonov, V. I. Perel', \textit{Sov. Phys. JETP Lett.} \textbf{13}, 467 (1971).
% 8
\bibitem{Hirsch1999}
J. E. Hirsch, \textit{Phys. Rev. Lett.} \textbf{83}, 1834 (1999).
% 9
\bibitem{Zhang2000}
S. Zhang, \textit{Phys. Rev. Lett.} \textbf{85}, 393 (2000). 
% 10
\bibitem{Sinova2004}
J. Sinova, D. Culcer, Q. Niu, N. A. Sinitsyn, T. Jungwirth, A. H. MacDonald, \textit{Phys. Rev. Lett.} \textbf{92}, 126603 (2004).
% 11
\bibitem{Valenzuela2006}
S. O. Valenzuela, M. Tinkham, \textit{Nature} \textbf{442}, 176-179 (2006).
% 12
\bibitem{Kato2004}
Y. K. Kato, R. C. Myers, A. C. Gossard, D. D. Awschalom, \textit{Science}  \textbf{306}, 1910-1913 (2004).
% 13
\bibitem{Wunderlich2005}
J. Wunderlich, B. Kaestner, J. Sinova, T. Jungwirth, \textit{Phys. Rev. Lett.} \textbf{94}, 047204 (2005).
% 14
\bibitem{Saitoh2006}
E. Saitoh, M. Ueda, H. Miyajima, G. Tatara, \textit{Appl. Phys. Lett.}  \textbf{88}, 182509 (2006).
% 15
\bibitem{Kimura2007}
T. Kimura, Y. Otani, T. Sato, S. Takahashi, S. Maekawa, \textit{Phys. Rev. Lett.} \textbf{98}, 156601 (2007).
% 16
\bibitem{Pai2012}
C.-F. Pai, L. Liu, H. W. Tseng, D. C. Ralph, R. A. Buhrman, \textit{Appl. Phys. Lett.} \textbf{101}, 122404 (2012).
% 17
\bibitem{Liu2012a}
L. Liu, C.-F. Pai, Y. Li, H. W. Tseng, D. C. Ralph, R. A. Buhrman, \textit{Science} \textbf{336}, 555-558 (2012).
% 18
\bibitem{Liu2012b}
L. Liu, O. J. Lee, T. J. Gudmundsen, D. C. Ralph, R. A. Buhrman, \textit{Phys. Rev. Lett.} \textbf{109}, 096602 (2012).
% 19
\bibitem{Liu2012c}
L. Liu, C.-F. Pai, D. C. Ralph, R. A. Buhrman, \textit{Phys. Rev. Lett.} \textbf{109}, 186602 (2012).
% 20
\bibitem{Manchon2019}
A. Manchon, J. \v{Z}elezn\'{y}, I. M. Miron, T. Jungwirth, J.Sinova, A. Thiaville, K. Garello, P. Gambardella, \textit{Rev. Mod. Phys.} \textbf{91}, 035004 (2019).
% 21
\bibitem{Zhu2018}
L. Zhu, D. C. Ralph, R. A. Buhrman, \textit{Phys. Rev. Applied} \textbf{10}, 031001(R) (2018).
% 22
\bibitem{Demasius2016}
K.-U. Demasius, T. Phung, W. Zhang, B. P. Hughes, S.-H. Yang, A. Kellock, W. Han, A. Pushp, S. S. P. Parkin, \textit{Nat. Commun.} \textbf{7}, 10644 (2016).
% 23
\bibitem{An2018a}
H. An, Y. Kanno, A. Asami, K. Ando, \textit{Phys. Rev. B} \textbf{98}, 014401 (2018).
% 24
\bibitem{An2018b}
H. An, T. Ohno, Y. Kanno, Y. Kageyama, Y. Monnai, H. Maki, J. Shi, K. Ando, \textit{Sci. Adv.} \textbf{4}: eaar2250 (2018).
% 25
\bibitem{Zhu2019a}
L. Zhu, L. Zhu, S. Shi, M. Sui, D. C. Ralph, R. A. Buhrman, \textit{Phys. Rev. Applied} \textbf{11}, 061004(R) (2019).
% 26
\bibitem{Zhu2019b}
L. Zhu, R. A. Buhrman, \textit{Phys. Rev. Applied} \textbf{12}, 051002(R) (2019).
% 27
\bibitem{Althammer2013}
M. Althammer, S. Meyer, H. Nakayama, M. Schreier, S. Altmannshofer, M. Weiler, H. Huebl, S. Gepr\"{a}gs, M. Opel, R. Gross, D. Meier, C. Klewe, T. Kuschel, J.-M. Schmalhorst, G. Reiss, L. Shen, A. Gupta, Y.-T. Chen, G. E. W. Bauer, E. Saitoh, S. T. B. Goennenwein, \textit{Phys. Rev. B} \textbf{87}, 224401 (2013).
% 28
\bibitem{Zhang2013}
W. Zhang, V. Vlaminck, J. E. Pearson, R. Divan, S. D. Bader, A. Hoffmann, \textit{Appl. Phys. Lett.} \textbf{103}, 242412 (2013).
% 29
\bibitem{Seki2008}
T. Seki, Y. Hasegawa, S. Mitani, S. Takahashi, H. Imamura, S. Maekawa, J. Nitta, K. Takanashi, \textit{Nat. Mater.} \textbf{7}, 125-129 (2008).
% 30
\bibitem{Rashba1960}
E. Rashba, \textit{Sov. Experimental details on thin film growth and magnetic characterization; Si metasurface fabrication; transmission, reflection, and dissymmetry measurements; and full-field electromagnetic simulations can be found in the Supporting Information. Phys. Solid State} \textbf{2}, 1109-1122 (1960).
% 31
\bibitem{Edelstein1990}
M. Edelstein, \textit{Solid State Commun.} \textbf{73}, 233 (1990).
% 32
\bibitem{Bychkov1984}
Y. A. Bychkov, E. I. Rasbha, \textit{Sov. Phys. JETP.} \textbf{39}, 66-69 (1984).
% 33
\bibitem{Burkov2004}
A. A. Burkov, A. S. N\'{u}\~{n}ez, A. H. MacDonald, \textit{Phys. Rev. B} \textbf{70}, 155308 (2004). 
% 34
\bibitem{Rojas-Sanchez2013}
J. C. Rojas-S\'{a}nchez, L. Vila, G. Desfonds, S. Gambarelli, J. P. Attane, J. M. De Teresa, C. Magen, A. Fert, \textit{Nat. Commun.} \textbf{4}, 2944 (2013).
% 35
\bibitem{Jungfleisch2016}
M. B. Jungfleisch, W. Zhang, J. Sklenar, W. Jiang, J. E. Pearson, J. B. Ketterson, A. Hoffmann, \textit{Phys. Rev. B} \textbf{93}, 224419 (2016).
% 36
\bibitem{LaShell1996}
S. LaShell, B. A. McDougall, E. Jensen, \textit{Phys. Rev. Lett.} \textbf{77}, 3419 (1996).
% 37
\bibitem{Nicolay2001}
G. Nicolay, F. Reinert, S. H\"{u}fner, P. Blaha, \textit{Phys. Rev. B} \textbf{65}, 033407 (2001).
% 38
\bibitem{Nechaev2009}
I.A. Nechaev, M. F. Jensen, E. D. L. Rienks, V. M. Silkin, P. M. Echenique, E. V. Chulkov, Ph. Hofmann, \textit{Phys. Rev. B} \textbf{80}, 113402 (2009).
% 39
\bibitem{Johnson1993}
M. Johnson, \textit{Phys. Rev. Lett.} \textbf{70}, 2142 (1993).
% 40
\bibitem{Ji2004}
Y. Ji, A. Hoffmann, J. S. Jiang, S. D. Bader, \textit{Appl. Phys. Lett.} \textbf{85}, 6218 (2004).
% 41
\bibitem{Ku2006}
J. Ku, J. Chang, h. Kim, J. Eom, \textit{Appl. Phys. Lett.} \textbf{88}, 172510 (2006).
% 42
\bibitem{Mihajlovic2009}
G. Mihajlovi\'{c}, J. E. Pearson, M. A. Garcia, S. D. Bader, A. Hoffmann, \textit{Phys. Rev. Lett.} \textbf{103}, 166601 (2009).
% 43
\bibitem{Seki2010}
T. Seki, I. Sugai, Y. Hasegawa, K. Takanashi, \textit{Solid State Commun.} \textbf{150}, 496-499 (2010).
% 44
\bibitem{Brangham2016}
J. T. Brangham, K.-Y. Meng, A. S. Yang, J. C. Gallagher, B. D. Esser, S. P. White, S. Yu, D. W. McComb, P. C. Hammel, F. Yang, \textit{Phys. Rev. B} \textbf{94}, 054418 (2016).
% 45
\bibitem{Tian2016}
45 D. Tian, C. Chen, H. Wang, X. Jin, \textit{Chin. Phys. B} \textbf{25}, 107201 (2016).
% 46
\bibitem{Chen2019}
C. Chen, D. Tian, H. Zhou, D. Hou, X. Jin, \textit{Phys. Rev. Lett.} \textbf{122}, 016804 (2019).
% 47
\bibitem{Li2019}
S. Li, K. Shen, K. Xia, \textit{Phys. Rev. B} \textbf{99}, 134427 (2019).
% 48
\bibitem{Bondarenko2013}
L. V. Bondarenko, D. V. Gruznev, A. A. Yakovlev, A. Y. Tupchaya, D. Usachov, O. Vilkov, A. Fedorov, D. V. Vyalikh, S. V. Eremeev, E. V. Chulkov, A. V. Zotov, A. A. Saranin, \textit{Sci. Rep.} \textbf{3}, 1826 (2013).
% 49
\bibitem{Shen2015}
H. Shen, D. Lu, B. VanSaders, J. J. Kan, H. Xu, E. E. Fullerton, Z. Liu, \textit{Phys. Rev. X} \textbf{5}, 021021 (2015).
% 50
\bibitem{Qian2021}
H. Qian, M. S. El Hadri, X. Wu, L. Chen, E. E. Fullerton, Z. Liu, \textit{Unpublished} (2021).
% 51
\bibitem{Supplemental} See the Supplemental Material for methods, high-resolution transmission electron microscopy image of ultrathin Au/Si multilayers, and fitting of the non-local resistance measured for 60-nm-thick Au
film.
% 52
\bibitem{Cherradi1989}
N. Cherradi, A. Audouard, G. Marchal, J. M. Broto, A. Fert, \textit{Phys. Rev. B}  \textbf{39}, 7424 (1989).
% 53
\bibitem{Abanin2009}
D. A. Abanin, A. V. Shytov, L. S. Levitov, B. I. Halperin, \textit{Phys. Rev. B} \textbf{79}, 035304 (2009).
% 54
\bibitem{Villamor2013}
E. Villamor, M. Isasa, L. E. Hueso, F. Casanova, \textit{Phys. Rev. B} \textbf{87}, 094417 (2013).
% 55
\bibitem{Liu2011}
L. Liu, T. Moriyama, D. C. Ralph, R. A. Buhrman, \textit{Phys. Rev. Lett.} \textbf{106}, 036601 (2011).
% 56
\bibitem{Isasa2015}
M. Isasa, E. Villamor, L. E. Hueso, M. Gradhand, F. Casanova, \textit{Phys. Rev. B} \textbf{91}, 024402 (2015).
% 57
\bibitem{Pai2015}
C.-F. Pai, Y. Ou, L. H. Vilela-Le\~{a}o, D. C. Ralph, R. A. Buhrman, \textit{Phys. Rev. B} \textbf{92}, 064426 (2015).
% 58
\bibitem{Sagasta2016}
E. Sagasta, Y. Omori, M. Isasa, M. Gradhand, L. E. Hueso, Y. Niimi, Y.C. Otani, F. Casanova, \textit{Phys. Rev. B} \textbf{94}, 060412(R) (2016).
% 59
\bibitem{Tanaka2008}
T. Tanaka, H. Kontani, M. Naito, T. Naito, D. S. Hirashima, K. Yamada, J. Inoue, \textit{Phys. Rev. B} \textbf{77}, 165117 (2008). 
% 60
\bibitem{Guo2009}
G. Y. Guo, \textit{J. Appl. Phys.} \textbf{105}, 07C701 (2009).
% 61
\bibitem{Yao2005}
Y. Yao, Z. Fang, \textit{Phys. Rev. Lett.} \textbf{95}, 156601 (2005).
% 62
\bibitem{Gu2010}
B. Gu, I. Sugai, T. Ziman, G. Y. Guo, N. Nagaosa, T. Seki, K. Takanashi, S. Maekawa, \textit{Phys. Rev. Lett.} \textbf{105}, 216401 (2010).
% 63
\bibitem{Guo2009}
G.-Y. Guo, S. Maekawa, N. Nagaosa, \textit{Phys. Rev. Lett.} \textbf{102}, 036401 (2009).
% 64
\bibitem{Zhou2015}
L. Zhou, V. L. Grigoryan, S. Maekawa, X. Wang, J. Xiao, \textit{Phys. Rev. B} \textbf{91}, 045407 (2015).
% 65
\bibitem{Kan2014}
J. J. Kan, \textit{Ph.D. Thesis}, UC San Diego, \textbf{2014}.
% 66
\bibitem{Ast2007}
C. R. Ast, J. Henk, A. Ernst, L. Moreschini, M. C. Falub, D. Pacil\'{e}, P. Bruno, K. Kern, M. Grioni, \textit{Phys. Rev. Lett.} \textbf{98}, 186807 (2007). \\
\end{thebibliography}

\begin{thebibliography}{2}


%1
\bibitem{Mihajlovic2009} G. Mihajlovi\'{c}, J. E. Pearson, M. A. Garcia, S. D. Bader, A. Hoffmann, \textit{Phys. Rev. Lett.} \textbf{103}, 166601 (2009).
\end{thebibliography}
\end{document}